\documentclass[12pt]{iopart}
\usepackage{graphicx}

\begin{document}

\title[Field-driven phase transitions in a quasi 2d quantum AFM]
{Field-driven phase transitions in a quasi-two-dimensional quantum
antiferromagnet}

\author{M. B. Stone$^{1,2,4}$, C. Broholm$^{2,3}$, D. H. Reich$^2$, P.
Schiffer$^4$, O. Tchernyshyov$^{2}$, P. Vorderwisch$^5$ and N.
Harrison$^6$}

\address{$^1$ Neutron Scattering Science Division, Oak Ridge National Laboratory, Oak Ridge, Tennessee 37831 USA}
\address{$^2$ Department of Physics and Astronomy, The Johns Hopkins University, Baltimore, Maryland 21218 USA}
\address{$^3$ National Institute of Standards and Technology, Gaithersburg, Maryland 20899 USA}
\address{$^4$ Department of Physics, Pennsylvania State University, University Park, Pennsylvania 16802 USA}
\address{$^5$ Hahn-Meitner Institut, D-14109 Berlin, Germany}
\address{$^6$ National High Magnetic Field Laboratory, LANL, Los Alamos, New Mexico 87545 USA}

\eads{\mailto{stonemb@ornl.gov}, \mailto{broholm@jhu.edu}}

\begin{abstract}
We report magnetic susceptibility, specific heat, and neutron
scattering measurements as a function of applied magnetic field and
temperature to characterize the $S=1/2$ quasi-two-dimensional
frustrated magnet piperazinium hexachlorodicuprate (PHCC). The
experiments reveal four distinct phases. At low
temperatures and fields the material forms a quantum paramagnet with
a 1 meV singlet triplet gap and a magnon bandwidth of 1.7 meV. The
singlet state involves multiple spin pairs some of which have
negative ground state bond energies. Increasing the field at low
temperatures induces three dimensional long range antiferromagnetic
order at 7.5 Tesla through a continuous phase transition that can be
described as magnon Bose-Einstein condensation. The phase transition
to a fully polarized ferromagnetic state occurs at 37 Tesla. The
ordered antiferromagnetic phase is surrounded by a renormalized
classical regime. The crossover to this phase from the quantum
paramagnet is marked by a distinct anomaly in the magnetic
susceptibility which coincides with closure of the finite
temperature singlet-triplet pseudo gap. The phase boundary between
the quantum paramagnet and the Bose-Einstein condensate features a
finite temperature minimum at $T=0.2$~K, which may be associated
with coupling to nuclear spin or lattice degrees of freedom close to
quantum criticality.
\end{abstract}
\submitto{New Journal of Physics}

\maketitle

\section{Introduction}

Molecular magnets form intricate quantum many body systems that are
amenable to experimental inquiry. Their low energy degrees of freedom
are quantum spins associated with transition metal ions such as
copper (spin-1/2) or nickel (spin-1). The interactions between these spins can be described by a Heisenberg
Hamiltonian of the form
\begin{equation}
    {\mathcal H} = \frac{1}{2}\sum_{{\mathbf r},{\mathbf d}}
    J_{d}{\mathbf S_{r}}\cdot {\mathbf S}_{{\mathbf r}+{\mathbf d}},
    \label{eq:heisenberg}
\end{equation}
Experiments on these materials can test existing theoretical
predictions and also reveal entirely new cooperative
phenomena\cite{hatfieldchapterdejonghreview,dejongh,broholmaeppli,broholmmagnetized,geiloreview}.

If the generic ground state of transition metal oxides is the
N\'{e}el antiferromagnet (AFM), it is singlet ground state magnetism for
molecular magnets. When molecular units have an even number of
spin-1/2 degrees of freedom they typically form a molecular singlet
ground state. If inter-molecular interactions are sufficiently weak
the result can be a quantum paramagnet. This is the state that
Landau famously envisioned as a likely outcome of antiferromagnetic
exchange interactions.\cite{Landau33} There are also indications
that geometrical frustration on lattices containing multiple
triangular plaquettes can suppress N\'{e}el order and/or produce
singlet ground state magnetism as is seen for example in the 3D
system $\mathrm{Gd_{3}Ga_{5}O_{12}}$ (GGG)
\cite{ramirezjapplphs1991,tsui1999} and possibly in
CuHpCl.\cite{stonecuhpclprb}

While magnets with an isolated singlet ground state generally have
no zero field phase transitions, there are at least two field
induced phase transitions that terminate in zero temperature quantum
critical points.  The lowest lying excitations in such systems
typically have spin $S = 1$, and thus the lower field critical point,
$H_{c1}$, occurs as
the $S_z=1$ triplet branch is field driven to degeneracy with the
singlet groundstate at a certain critical wave vector. The system
becomes magnetized even in the zero temperature limit for fields
above $H_{c1}$. As the field is increased further, the system is
ultimately ferromagnetically polarized at a second QCP characterized by an upper critical field, $H_{c2}$ where the field energy
overwhelms the spin-spin interactions.  Above this field, a second
spin-gap should open and increase linearly with applied field above
$H_{c2}$.

What happens as a function of field and temperature between these
critical points depends critically on the inter-molecular
connectivity. Three-dimensional systems such as
$\mathrm{Cs_{3}Cr_{2}Br_{9}}$\cite{leuenbergerprb1985} and
$\mathrm{TlCuCl_{3}}$
\cite{oosawaphysicab2001,oosawajphys1999,cavadiniprb63,cavadininature63}
exhibit a phase transition to long range antiferromagnetic order
both as a function of applied field and applied pressure. For
quasi-one-dimensional connectivity the high field phase should be an
extended quantum critical phase. However, systems that have been examined thus far, such as
$\mathrm{Cu(NO_{3})_{2}\cdot 2.5 H_{2}O}$
\cite{stonecoppernitrateprb,gxuprl2000}, and
$\mathrm{Cu(L-aspartato)(H_{2}O)_{2}}$\cite{calvoprb1999} exhibit long range N\'{e}el order indicating
that they may not be as one-dimensional as originally envisioned.
Another complication can be multiple molecules per unit cell which lead
to staggered fields that break translation symmetry elements and
preclude a critical phase. Several new spin ladder systems have
recently been identified that may enable access to a critical phase
above $H_{c1}$.\cite{quinoxaline,landeenew}

Two-dimensional (2D) systems are of particular interest since they
are at the lower critical dimension for N\'{e}el as the parent
compounds to high temperature superconductors.
$\mathrm{BaCuSi_{2}O_{6}}$ for example features spin-1/2 pairs
coupled on a two dimensional square lattice.\cite{sasagoprb1997} An
anomaly in the critical exponents and in the phase boundary at low
temperatures has been interpreted as the result of a dimensional
crossover from three to two dimensions close to quantum
criticality.\cite{sebastianpreprint} In quasi-two dimensional
$\mathrm{SrCu_{2}(BO_{3})_{2}}$ competing interactions play a
central role. Its unique geometry frustrates inter-dimer
interactions, yielding effectively decoupled dimers
\cite{kageyamaprl2000}. While the singlet ground state of the
material is known exactly, the three finite-magnetization plateaus
and the critical phases that surround them are more challenging to
understand. The purely organic quantum AFM F$_2$PNNO forms a
quasi-2D network of spin dimers with a singlet ground state and a 1
meV gap to magnetic excitations.\cite{pureorgo} The phase diagram
includes a magnetization plateau that extends from $\mu_0H=15$ T to
$\mu_0H=25$ T as well as regions with continuously varying
magnetization that may indicate unique 2D quantum critical phases.
Neutron scattering is required to disentangle these complex and
unfamiliar high field phases. Unfortunately few of these phases can
be accessed with current high field neutron scattering facilities,
which are limited to fields below 17.5 Tesla.

\begin{figure}[t]
\centering\includegraphics[scale=0.5]{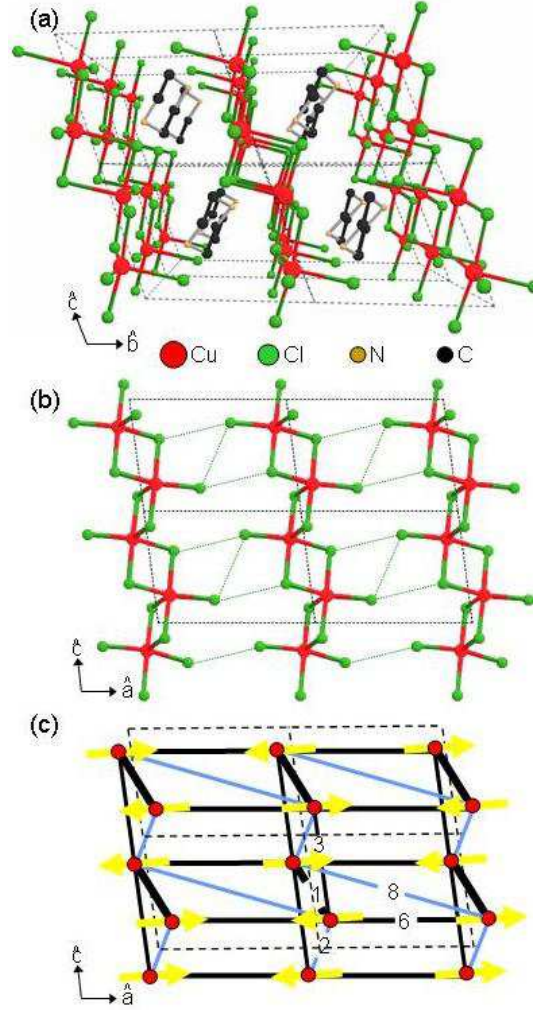}
\caption{\label{fig:phccstructure}Crystal structure of piperazinium
hexachlorodicuprate (PHCC), $\rm (C_{4}H_{12}N_{2})(Cu_{2}Cl_{6})$.
(a) View along the {\bf a} axis, showing well-separated Cu-Cl
planes. (b) View of a single Cu-Cl plane.  Dotted lines between Cl
sites show potential halide-halide contacts. (c) Second view of a
single Cu-Cl plane, showing only the interacting Cu$^{2+}$ spin
sites (solid circles). Significant antiferromagnetic spin-spin
interactions are depicted as light blue (\emph{frustrated}) and
black (\emph{satisfied}) lines with the same bond numbering used in
\cite{stonephccprb01}. Line thickness is proportional to the product
of exchange constant and spin-spin correlation function,
$|J_{\mathbf{d}}\langle
{\mathbf{S}}_{\mathbf{0}}\cdot{\mathbf{S}}_{\mathbf{d}}\rangle|$.
Vectors represent the ordered spin structure determined by elastic
neutron scattering at $T=1.65$ K and $\mu_0 H=13.7$ T. Dashed lines
are the $\bf{a-c}$ plane unit cell with room temperature lattice
constants $a = 7.984(4)$ \AA, $b = 7.054(4)$ \AA, $c = 6.104(3)$
\AA, and $\alpha = 111.23(8)^{\circ}$, $\beta = 99.95(9)^{\circ}$,
and $\gamma = 81.26(7)^{\circ}$, space group $P\bar{1}$, from
\cite{battagliastruct}.}
\end{figure}

The material examined here, piperazinium hexachlorodicuprate (PHCC),
is a highly two dimensional quantum paramagnet with a lower critical
field of 7.5 T that is well within reach of current high field
neutron scattering and an upper critical field of 37 T that can be
reached in pulsed field magnetization measurements. The large
difference between the critical fields indicates strong inter-dimer
interactions and a potential for novel, strongly correlated
two-dimensional physics. PHCC,
$\mathrm{(Cu_{4}H_{12}N_{2})Cu_{2}Cl_{6}}$, consists of Cu$^{2+}$
spin-$\frac{1}{2}$ sites interacting predominantly within the
crystallographic {\bf
a-c}-planes\cite{stonephccprb01,stonephccnature}, as illustrated in
figure~\ref{fig:phccstructure}.  Prior zero-field inelastic neutron
scattering measurements probed the spin dynamics of the
low-temperature quantum paramagnetic (QP) phase, and established the
strongly 2D nature of magnetism in PHCC.  The spin gap $\Delta =
1.0$ meV, and the magnetic excitations show dispersion with an $1.8$
meV bandwidth in the $(h0l)$ plane. There is less than 0.2 meV dispersion
perpendicular to that plane, indicating a highly two dimensional spin
system. The spin-spin interactions in the {\bf a-c}-plane have the
connectivity of a frustrated orthogonal bilayer, as shown in
figure~\ref{fig:phccstructure}(c). Bonds 6 and 3 form two extended
2D lattices linked by Bond 1, which thus is $J_{\perp}$ in the
bilayer description.  Bonds 2 and 8 provide additional frustrated
interlayer interactions.

In the current study, we employed inelastic and elastic neutron
scattering as well as specific heat and magnetic susceptibility
measurements to characterize the low-temperature phases of PHCC.
Unlike 3D coupled-dimer systems, such as $\mathrm{TlCuCl_{3}}$, we
find a significant separation between the phase boundary to long
range order (LRO) and the crossover line delineating the high-field
boundary of the QP regime.  This allows access to a disordered
renormalized classical (RC-2D) regime near the lower QCP. Specific
heat measurements in the disordered low-field QP phase agree well
with series expansions from a quantum bilayer model.  Combining
elastic neutron scattering and field dependent specific heat results
allows a determination of the critical exponents associated with the
transition to LRO. We have previously reported a reentrant phase
transition at low temperature near $H_{c1}$ wherein the magnetic LRO
can melt into the QP phase as the temperature is either increased or
decreased \cite{phccprl1}. We have proposed that coupling between
other degrees of freedom is important near the QCP, and thus can
have a non-negligible effect on the phase diagram in this regime. We
also have recently examined the LRO transition and found good
agreement with a representation of the phase boundary away from the
reentrant phase transition in terms of magnon Bose-Einstein
condensation. In this communication, we provide further experimental
and theoretical information regarding these phenomena, including
field dependent neutron scattering, magnetization, and specific heat
data, and a more detailed theoretical discussion of magnon
Bose-Einstein condensation in PHCC.

\section{Experimental techniques}
Single crystal  samples of deuterated and hydrogenous PHCC were
prepared via previously described methods \cite{stonephccprb01}.
Magnetic susceptibility measurements were performed at the National
High Magnetic Field Laboratory pulsed field facility at the Los
Alamos National Laboratory using a compensated coil susceptometer.
The sample environment consisted of a pumped He$^{3}$ cryostat
located in the bore of a 50 Tesla pulsed field resistive magnet.
Measurements were performed with the magnetic field along the {\bf
b}-, {\bf a}$^{*}$- and {\bf c}- axes on hydrogenous samples with
masses between $m = 0.7$ mg and $m = 1.4$ mg.  An empty coil
measurement was performed for each individual pulsed field
measurement of the sample as a direct measure of the
sample-independent background.  Specific heat measurements were
performed as a function of magnetic field, $\mu_0 H \leq 14$ T, and
temperature, $T \geq 1.8$ K, on a deuterated single crystal sample,
$m = 6.76$ mg, with ${\bf H} \parallel$ {\bf a}$^{*}$ using a commercial
calorimeter (Quantum Design PPMS).

Neutron scattering measurements were performed using the FLEX triple
axis spectrometer at the cold neutron facility of the Hahn-Meitner
Institut in Berlin, Germany.  The sample was mounted on an aluminum
holder and attached to the cold finger of a dilution refrigerator
within a split coil $\mu_0 H=14.2$ T superconducting magnet. A magnetic
field in excess of the superconducting critical field for aluminum
was always present to ensure good thermal contact to the sample. The
sample consisted of two deuterated single crystals with a total mass
of $m=1.75$ g coaligned within 0.5$^{\circ}$.  The sample was
oriented in the $(h0l)$ scattering plane with ${\bf H}
\parallel$ {\bf b}. Measurements were performed using a vertically
focusing pyrolytic graphite (PG(002)) monochromator followed by a
$60^{\prime}$ horizontal collimator. Inelastic measurements used a
liquid nitrogen cooled Be filter before the sample and a fixed final
energy of $E_{f}=3$ meV. A horizontally focusing PG(002) analyzer
was used for inelastic measurements. Elastic measurements employed a
flat PG(002) analyzer with $60^{\prime}$ collimation before and
after the analyzer. Measurements characterizing the temperature and
magnetic field dependence of Bragg peaks used a liquid nitrogen
cooled Be filter before the sample and neutron energy
$E_{i}=E_{f}=2.5$ meV. Measurements of magnetic and nuclear Bragg
peaks for determining the ordered magnetic structure were performed
using the same collimation with a room temperature PG filter before
the sample and $E_{i}=E_{f}=14.7$ meV with a vertically focusing
PG(004) monochromator and a flat PG(002) analyzer.

\section{Experimental results}
\subsection{Susceptibility and magnetization}
\begin{figure}[t]
\centering\includegraphics[scale=0.70]{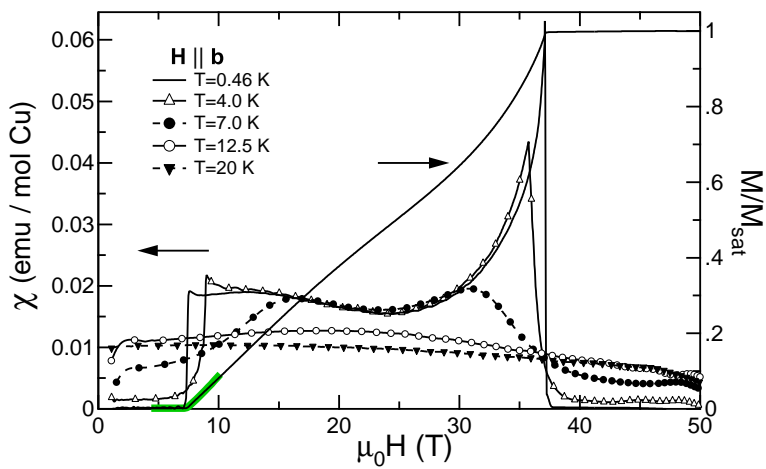}
\caption{\label{fig:phccprb2chiofhparb}
Magnetic susceptibility $\chi$ (left axis) and normalized magnetization
$M/M_{sat}$ (right axis) ($T=0.46$ K only) of PHCC vs. magnetic field
for ${\bf H}\parallel {\mathbf b}$ at different temperatures.  Markers are
plotted for every thirtieth data point.  The green line in the vicinity
of $H_{c1}$ is a fit to $M(H)$, as described in the text.}
\end{figure}

Pulsed field measurements of the magnetic susceptibility $\chi(H)$
up to $\mu_0 H=50$ T with ${\bf H} \parallel{\mathbf b}$ are shown in
figure~\ref{fig:phccprb2chiofhparb}.  For $T\leq 4$ K, the
susceptibility has two sharp features.  The lower field feature
marks closure of the spin gap while the upper feature indicates the
field where $M(H)$ saturates, and the system enters the fully
polarized (FP) regime.  $M(H)$ at $T = 0.46$ K, obtained by
integrating the corresponding $\chi(H)$ data, is shown as a solid
line in figure~\ref{fig:phccprb2chiofhparb}.  At low temperatures,
these features in $\chi(H)$ indicate the location of the critical
fields $\mu_0 H_{c1} \approx 7.5$ T and $\mu_0 H_{c2} \approx 37$ T.
We note that although there is a local minimum in $\chi(H)$ at
$\mu_0 H = 23.9$ T near $H_m = \frac{1}{2}(H_{c1}+H_{c2})$, and
hence a point of inflection in the
magnetization, there is no indication of fractional magnetization
plateaus for $T \ge 0.46$ K.

\begin{figure}[tb]
\centering\includegraphics[scale=0.65]{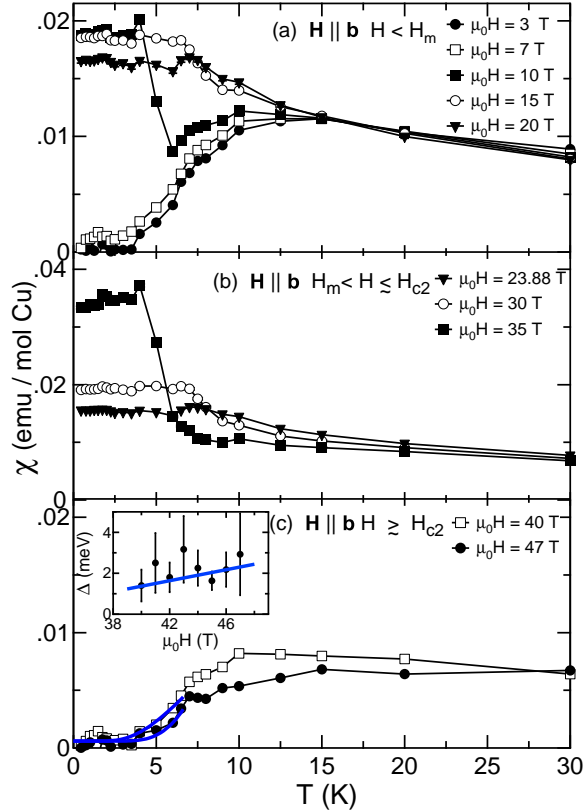}
\caption{\label{fig:phccprb2chioft} Magnetic susceptibility of PHCC
as a function of temperature for ${\bf H} \parallel {\mathbf b}$.
(a) $\mu_0 H \le 20$ T, (b) 23 T $\le \mu_0 H < \mu_0 H_{c2}$, (c)
$H>H_{c2}$. Data were obtained by combining all individual pulsed
field measurements and averaging over the range $\mu_0 (\Delta  H) =
0.5$ T for each magnetic field shown.  Lines connecting data points
are guides to the eye.  Curves in (c) for $T < 8 $ K are fits
described in the text that yield the field dependence of the gap
above $H_{c2}$ shown in the inset to (c).  The solid line in the
inset to (c) has the slope $g\mu_B H$.}
\end{figure}

As $T$ increases, the two features in $\chi(H)$ broaden, and above
$T = 9$ K, they merge into a single rounded maximum that moves to
lower fields as the temperature is further increased.  For $T > 20$
K, $\chi(H)$ decreases monotonically with $H$.  Despite the thermal
broadening above $T=5$ K, several regimes may be identified from
$\chi(H,T)$.  These are most clearly seen in figure
\ref{fig:phccprb2chioft}, which shows $\chi(H,T)$ at fixed field.
The rounded peak and exponential decrease in $\chi(T)$ for low
magnetic fields is characteristic of a gapped antiferromagnet and is
similar to prior results obtained at $H=0$ for PHCC
\cite{stonephccprb01,daoudprb1986,battagliastruct}.  As the field
increases, the peak in $\chi(T)$ shifts to lower $T$, and the
low-temperature activated behavior vanishes with the spin-gap.  As
shown for $\mu_0 H = 10$ T in figure \ref{fig:phccprb2chioft}(a),
for fields $\mu_0 H_{c1} < \mu_0 H < 12$ T the rounded maximum that
signals the onset of 2D antiferromagnetic spin correlations leading
to the gapped state is still visible (near $T = 12$ K for $\mu_0 H =
10$ T), but there is then a sharp rise in $\chi(T)$ as the system
enters the strongly magnetizable regime at lower $T$. As shown in
figures~\ref{fig:phccprb2chioft}(a) and~\ref{fig:phccprb2chioft}(b)
the boundary of this regime continues to move to higher $T$ for
$\mu_0 H < $ 25 T, before moving back to lower $T$ as $H$ approaches
$H_{c2}$. In the FP region above $H_{c2}$  activated behavior
returns, as shown for $\mu_0 H=40$ T and $\mu_0 H=47$ T in
figure~\ref{fig:phccprb2chioft}(c).  This indicates a second
spin-gap regime.  Note that as the field increases above $H_{c2}$,
the onset of the activated susceptibility moves to higher $T$,
indicating that this second spin gap increases with field. An
overview of $\chi(H,T)$ is given in the color contour plot in figure
\ref{fig:phccprb2phasediag}(a) and will be discussed in more detail
below.

\begin{figure}[tb]
\centering\includegraphics[scale=0.67]{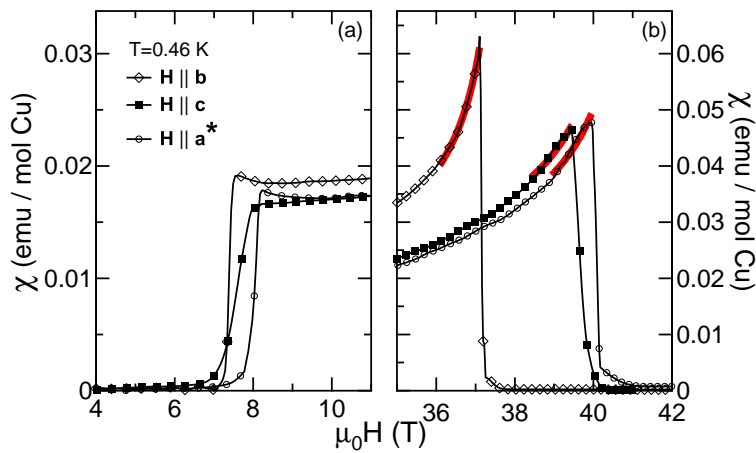}
\caption{\label{fig:phccprb2chithreeorient}
Magnetic susceptibility of PHCC at $T=0.46$ K for
${\bf H} \parallel {\mathbf b}$, ${\bf H} \parallel {\mathbf c}$, and
${\bf H} \parallel {\mathbf a^{*}}$ in the vicinity of (a)
$H_{c1}$ and (b) $H_{c2}$.  Data obtained for different
sample masses were normalized to the saturation magnetization
for ${\bf H} \parallel{\mathbf b}$.  Solid red lines are fits
in the vicinity of the upper critical fields as described in
the text.  Markers are plotted every seventh datapoint.}
\end{figure}

Low-temperature pulsed field susceptibility measurements were also
performed at $T = 0.46$ K for
${\bf H} \parallel \mathbf{c}$ and ${\bf H} \parallel \mathbf{a^{*}}$.
These are shown together with the ${\bf H} \parallel \mathbf{b}$ data
near $H_{c1}$ and $H_{c2}$ in figure~\ref{fig:phccprb2chithreeorient}.
Although the overall shape of  $\chi(H)$ is similar in all three
orientations, the differences in the onset fields shown in
figure~\ref{fig:phccprb2chithreeorient} indicate
that the critical fields are orientation-dependent.
Because the ratio of the critical fields is nearly the same
for each of the three single crystal orientations,
[$H_{c1}({\mathbf b})/H_{c2}({\mathbf b}) = $ 0.203(3),
$H_{c1}({\mathbf c})/H_{c2}({\mathbf c}) = $ 0.203(3),
and $H_{c1}({\mathbf  a}^{*})/H_{c2}({\mathbf a}^{*}) = $ 0.205(3)],
differences in the magnitude of the critical fields are likely due to
$g$-factor anisotropy associated with the coordination of the
Cu$^{2+}$ ions in PHCC.

\subsection{Specific heat}
\begin{figure}[t]
\centering\includegraphics[scale=0.45]{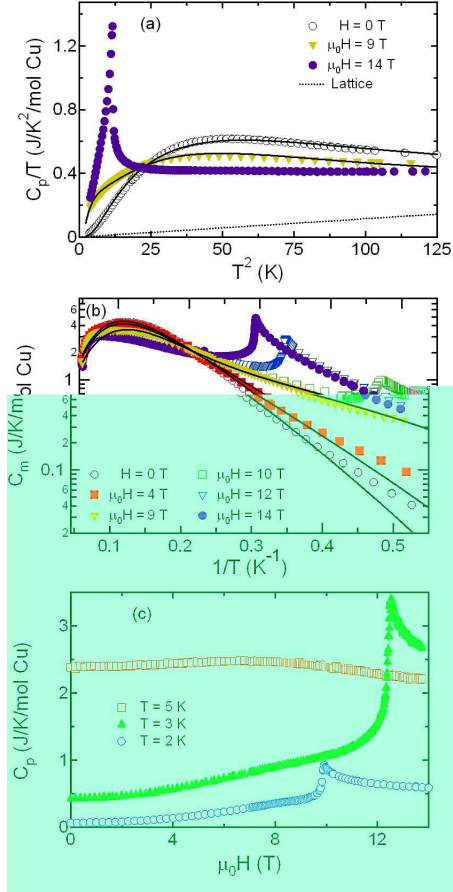}
\caption{\label{fig:phccprb2cp1} Specific heat as a function of
temperature and magnetic field for deuterated PHCC with ${\bf H}
\parallel {\mathbf a}^{*}$. (a) Total specific heat $C_p/T$ vs.
$T^2$ at fixed $H$.  Every tenth data point is plotted for $\mu_0 H
= 14$ T.  Calculated lattice contribution is shown as a dashed line.
(b) Magnetic specific heat $C_{m} = C_{p} - C_{\mathrm{lattice}}$
vs. $1/T$. Solid lines for $\mu_0 H < 10$ T in (a) and (b) are the
result of a global fit to the frustrated bilayer model discussed in
the text. (c)  Isothermal specific heat.}
\end{figure}

Measurements of the specific heat as a function of temperature and
magnetic field are shown in figure~\ref{fig:phccprb2cp1}. Prior
measurements were performed on protonated samples in the same
orientation, ${\bf H} \parallel {\mathbf a}^{*}$, but were restricted to $T
\leq 2$ K and $\mu_0 H \leq 9$ T.  The total specific heat $C_{p}$
is shown for representative fields in
figure~\ref{fig:phccprb2cp1}(a). In zero field, $C_p$ shows a broad
rounded maximum at $T\approx 7$ K, with an exponentially activated
region at lower $T$ associated with the formation of the spin-gap.
Above $H_{c1}$ the form of the high-temperature specific heat
changes, and as shown for $\mu_0 H = 14$ T, a lambda-type anomaly
can be seen that is associated with the phase transition to magnetic
LRO at $T = T_{N}$.  In figure~\ref{fig:phccprb2cp1}(b) we plot the
magnetic portion of the specific heat, $C_{m} = C_{p} -
C_{\mathrm{lattice}}$, where $C_{\mathrm{lattice}} \propto T^3$ is
the lattice/phonon contribution to the specific heat determined by
comparison of the data to the bilayer model discussed later in the
text.  This lattice contribution is shown as a dashed line in
figure~\ref{fig:phccprb2cp1}(a). Figure~\ref{fig:phccprb2cp1}(b)
shows the changes in the low-field activated behavior of $C_m$ as
$H$ increases and the spin-gap closes, as well as the increase in
magnitude of the lambda peaks as $H$ increases above $H_N$.  The
area under the peak is greater for larger applied fields indicating
that the $T=0$ ordered moment increases with magnetic field.  The
field dependence of $C_{p}$ is plotted in
figure~\ref{fig:phccprb2cp1}(c) for selected temperatures.  We observe
a sharp peak at the transition to LRO at $H = H_N$, which has the
same field and temperature dependence as that seen in $C_p(T)$ at
fixed $H$.

\subsection{Elastic neutron scattering}
Neutron diffraction was used to explore the nature of the LRO
transition indicated by heat capacity data. Field and temperature
dependent antiferromagnetic Bragg peaks were observed at wave vector
transfer ${\bf Q}=(\frac{(2n+1)}{2}$, $0$, $\frac{(2m+1)}{2})$ for
integer $n$ and $m$ throughout the $(h0l)$ plane, corresponding to a
magnetic structure with a doubling of the unit cell along the {\bf
a}- and {\bf c}-axes.

\begin{figure}[tb]
\centering\includegraphics[scale=0.62]{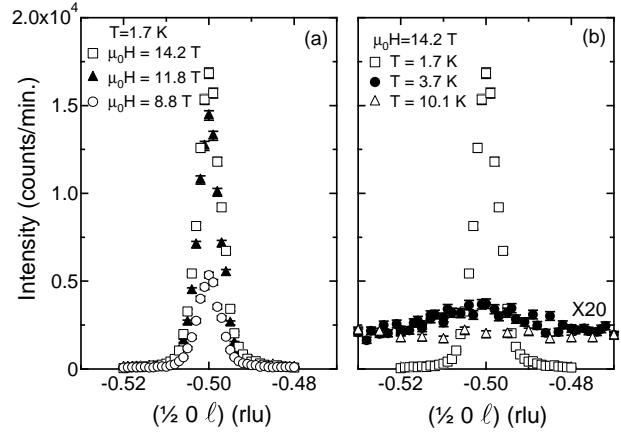}
\caption{\label{fig:phccprb2braggoverview} Wave-vector dependent
elastic scattering from PHCC as measured through the $(\frac{1}{2} 0
\bar{\frac{1}{2}})$ Bragg peak along the $(\frac{1}{2} 0 l)$
direction.  (a) Magnetic field, ${\bf H} \parallel {\mathbf b}$, dependence
at $T=1.7$ K. (b) Temperature dependence at $\mu_0 H=14.2$ T.}
\end{figure}

Figure~\ref{fig:phccprb2braggoverview} depicts a wave-vector scan
through the $(\frac{1}{2} 0 \bar{\frac{1}{2}})$ Bragg peak as a
function of (a) magnetic field and (b) temperature.  The scattering
intensity at $T=1.7$ K increases as a function of applied field.  At
$\mu_0 H=14.2$ T (figure~\ref{fig:phccprb2braggoverview}(b)) there
is only a flat background at $T=10.1$ K; however, for $T=3.7$ K the
broad maximum indicates short range AFM correlations. At $T=1.7$ K
the resolution limited peak indicates long range antiferromagnetic
order. Scans along the $(h 0 \bar{\frac{1}{2}})$ and $(\zeta 0
\bar{\zeta})$ directions were also resolution limited. Magnetic
Bragg peaks were found at locations that were indistinguishable from
the nuclear half-wavelength, $\lambda/2$, Bragg peaks, which were
observed upon removal of the Be filter to admit $\lambda/2$ neutrons
from the PG(004) monochromator Bragg reflection. No indication of
incommensurate order was found along the $(h 0 \bar{\frac{1}{2}})$
and $(\frac{1}{2} 0 l)$ directions through the entire Brillouin zone
at $T=1.9$ K and $\mu_0 H=11.8$ T.

\begin{figure}[tb]
\centering\includegraphics[scale=0.85]{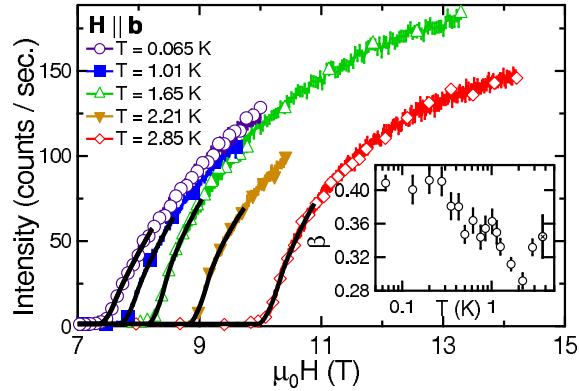}
\caption{\label{fig:phccprb2fieldsweeps} Magnetic field, ${\bf H}
\parallel {\mathbf b}$, dependence of elastic scattering intensity
of the $(\frac{1}{2} 0 \bar{\frac{1}{2}})$ Bragg peak at select
temperatures.  Temperatures presented increase according to the
figure legend from left to right.  Solid lines are fits described in
the text over the region $\mu_0 H < \mu_0 H_N + 0.75$ T.   Inset:
magnetic order parameter critical exponent, $\beta$, determined by
fitting a power law $I(H)\propto(H-H_N)^{2\beta}$ to magnetic field
and temperature dependent scattering from the $(\frac{1}{2} 0
\bar{\frac{1}{2}})$ antiferromagnetic Bragg peak.  The Data point at
$T = 3.69$ K corresponds to the temperature scan shown in
figure~\ref{fig:phccprb2tempsweeps}.}
\end{figure}

\begin{figure}[tb]
\centering\includegraphics[scale=0.85]{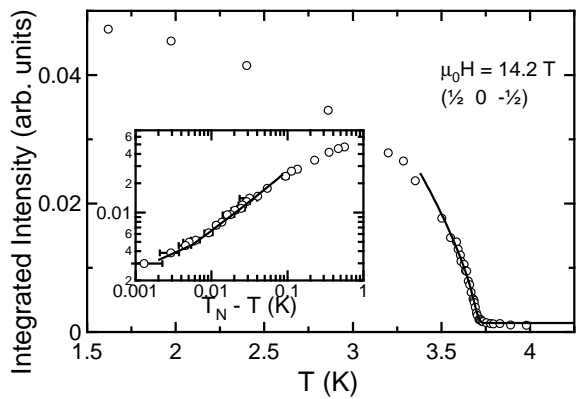}
\caption{\label{fig:phccprb2tempsweeps} Temperature dependence of
$(\frac{1}{2} 0 \bar{\frac{1}{2}})$ antiferromagnetic Bragg peak
elastic scattering intensity for PHCC at $\mu_0 H=14.2$ T with ${\bf
H} \parallel {\mathbf b}$.  Data points were obtained from the
integrated intensities of rocking curves measured at each
temperature. Inset: Log-Log plot of integrated intensity as a
function of the difference between the determined N\'{e}el
temperature, $T_{N}=3.705(3)$ K, and the sample temperature.  Solid
lines are fits as described in text over the range $T>T_{N}-0.33$ K.
}
\end{figure}

To characterize the onset of antiferromagnetic order, the
$(\frac{1}{2}0\bar{\frac{1}{2}})$ Bragg peak was measured as a
function of magnetic field at sixteen temperatures for $0.065\leq T
\leq 2.85$ K.  Data were obtained by continuously sweeping the
magnetic field ($\mu_0dH/dt=0.01$ or $0.03$ T/min.) while
simultaneously counting diffracted neutrons over a fixed time
interval of 0.5 minutes per point. At T=0.06 K the measured field
trace was indistinguishable for the slower and faster rates.  The
results are shown for select temperatures in
figure~\ref{fig:phccprb2fieldsweeps}. The temperature dependence of
the $(\frac{1}{2} 0 \bar{\frac{1}{2}})$ Bragg peak at $\mu_0 H=14.2$ T was
measured by integrating over a series of rocking curves at 37
temperatures as shown in figure~\ref{fig:phccprb2tempsweeps}.  The
intensity increases continuously upon cooling below a distinct
critical temperature as expected for a second order phase
transition.

\begin{figure}[t]
\centering\includegraphics[scale=0.65]{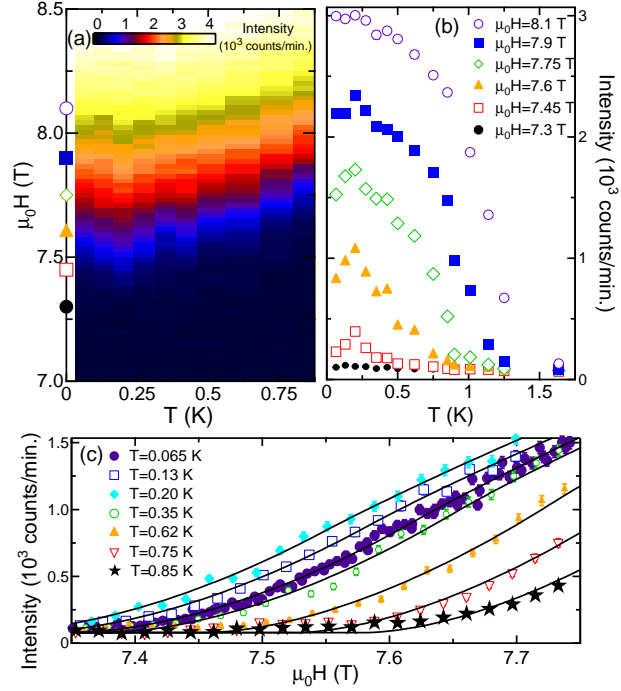}
\caption{\label{fig:reentrantplot} Reentrant behavior of
low-temperature LRO phase as a function of magnetic field, ${\bf
H}\parallel {\mathbf b}$, and temperature.  (a) Elastic scattering
intensity of the $(\frac{1}{2} 0 \bar{\frac{1}{2}})$
antiferromagnetic Bragg peak as a function of applied magnetic field
and temperature.  False color image obtained by combining all
individual magnetic field order parameter scans.  (b) Elastic
scattering intensity of the $(\frac{1}{2} 0 \bar{\frac{1}{2}})$
antiferromagnetic Bragg peak as a function of temperature.  Data
were obtained by combining all of the individual magnetic field
order parameter scans and averaging over $\mu_0 \Delta H = 0.1 $ T
for each magnetic field shown.  The symbols on the vertical axis in
(a) correspond to constant temperature data shown in (b).  (c)
Magnetic field order parameter scans of the $(\frac{1}{2} 0
\bar{\frac{1}{2}})$ antiferromagnetic Bragg peak performed in the
region of the reentrant phase transition.  Solid lines are fits
described in the text.}
\end{figure}

Figures~\ref{fig:reentrantplot}(a) and ~\ref{fig:reentrantplot}(c)
show low-temperature
magnetic field sweeps of the $(\frac{1}{2} 0 \bar{\frac{1}{2}})$
magnetic Bragg intensity plotted as a contour plot and as intensity
versus field in the vicinity of the critical field
for the transition to N\'{e}el order,
which we denote by $H_{N}$. The data indicate that the ordered phase
is thermally reentrant for $\mu_0 H\approx 7.5$ Tesla. Upon heating from
0.065 K, $H_{N}$ initially decreases with increasing temperature.
However, there is an apparent minimum in the critical field curve at
$T=0.20$ K, above which temperature $H_{N}$ increases on heating.
The same data are plotted as a function of temperature at fixed
fields in figure~\ref{fig:reentrantplot}(b). These data show a peak
in the scattering intensity at $T=0.20$ K for fields between
$\mu_0 H\approx 7.45$ T and $\mu_0 H\approx 7.9$ T.  The peak location remains
at $T\approx 0.2$ K until vanishing for fields above $\mu_0 H\approx 8.1$
T.

\subsection{Inelastic neutron scattering}
\begin{figure}[t]
\centering\includegraphics[scale=0.77]{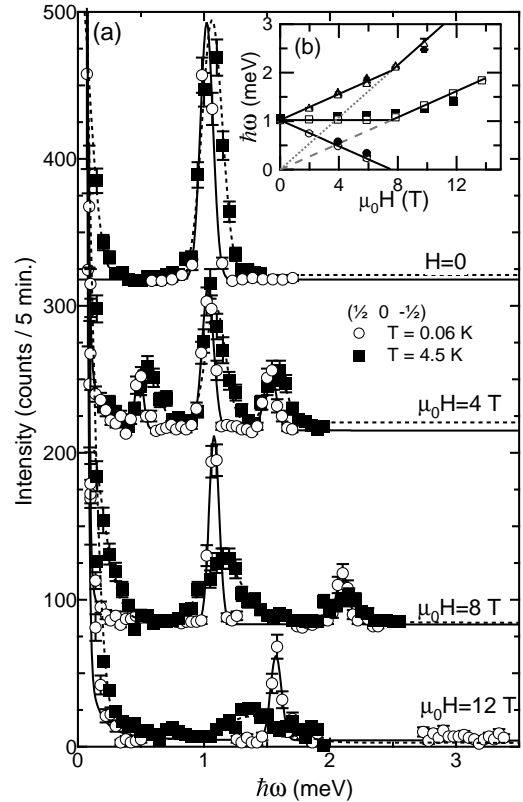}
\caption{\label{fig:phccprb2inelaslowt} (a)  Constant wave-vector
$(\frac{1}{2} 0 \bar{\frac{1}{2}})$ inelastic neutron scattering
from PHCC at $T=0.06$ K (open symbols) and $T=4.5$ K (solid symbols)
for different values of applied magnetic field, ${\bf H} \parallel {\mathbf
b}$.  Solid(dashed) lines are fits to Gaussians for the low(high)
temperature data as described in the text.  Data are offset
vertically for clarity.  (b) Peak locations derived from
 data such as that shown in
(a) as a function of applied magnetic field.  Solid lines are a
fit to the $T=0.06$ K data as described in text.  Dashed lines
extend these lines to the origin.}
\end{figure}

Inelastic neutron scattering was used to probe spin-dynamics as a
function of magnetic field and temperature. In particular, constant
wave vector scans at ${\bf Q} =(\frac{1}{2} 0 \bar{\frac{1}{2}})$  were
performed as a function of magnetic field at $T=0.06$ K and $T=4.5$
K.  A portion of these data is shown in
figure~\ref{fig:phccprb2inelaslowt}.  The low-temperature scan at
$H=0$ T features  a resolution limited excitation at $\hbar\omega
\approx 1$ meV, consistent with the previous zero field
measurements \cite{stonephccprb01}. In an applied field this
excitation splits into three components indicating a zero field
triplet 1 meV above a singlet ground state. The intensity
distribution is consistent with the upper and lower peaks resulting
from transitions to $S_{z}=\pm 1$ states while the central highest
intensity peak corresponds to a transition to an $S_{z}=0$ state.
The lower energy, $S_{z}=+1$, peak is observable until $\mu_0 H\approx 6$
T, where it is obscured by elastic nuclear incoherent scattering.
Both remaining gapped excitations increase in energy continuously as
the field is increased above $H_{c1}$.
The higher energy excitation increases in
energy and decreases in intensity at a faster rate than the
excitation which evolves out of the $S_{z}=0$ component of the
$H<H_{c1}$ triplet. There are also subtle differences in the
temperature and field dependence of the width of the excitations,
which we shall return to later.

\section{Discussion}
\subsection{Susceptibility and magnetization}
\begin{figure}[t]
\centering\includegraphics[scale=0.45]{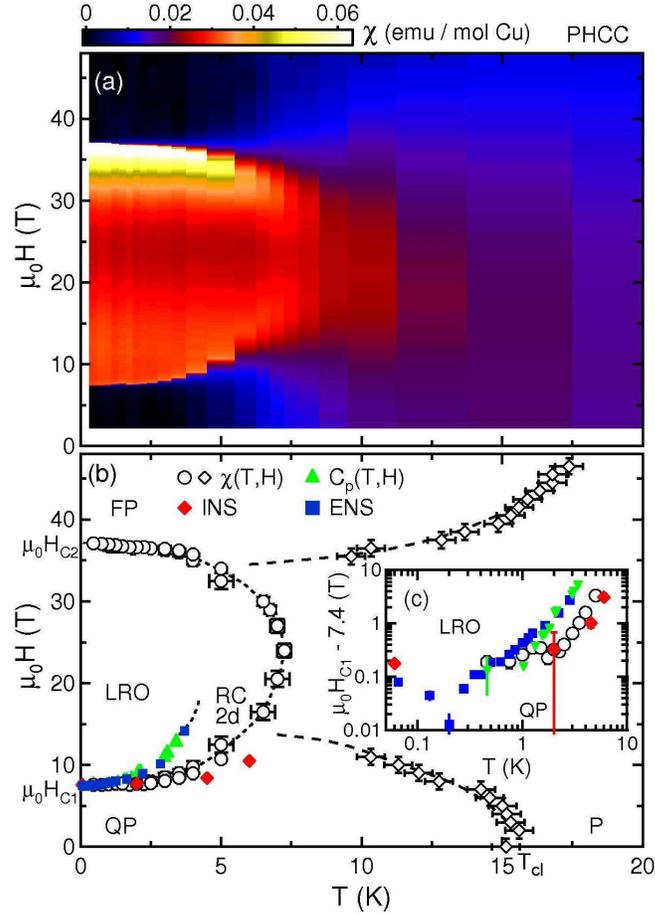}
\caption{\label{fig:phccprb2phasediag} (a) Color contour map of
$\chi(H,T)$ for PHCC.  (b) Magnetic field versus temperature phase
diagram of PHCC.  $H_{c1}$ and $H_{c2}$ are zero-temperature QCPs
described in the text.  Data points were obtained from pulsed field
susceptibility ($\chi (T,H)$), inelastic neutron scattering (INS),
elastic neutron scattering (ENS), and specific heat measurements
($C_P (T,H)$).  Specific heat measurements were performed with ${\bf
H}\parallel {\mathbf a}^{*}$; all other measurements were made with
${\bf H}\parallel {\mathbf b}$. Values from $C_P(T,H)$ have been
renormalized as discussed in the text.  LRO indicates a long range
ordered phase; QP is a quasi-2D quantum paramagnetic phase; FP
corresponds to a ferromagnetic polarized phase; RC-2D indicates a
quasi-2D renormalized classical phase; and P corresponds to the
semi-classical paramagnetic region.  Dashed lines are guides to the
eye.  Open diamonds represent the rounded peak locations in
$\chi(T)$ delineating a higher temperature classical paramagnetic
regime from the correlated low-temperature regimes, QP and FP.  Open
circles represent locations of sharp features in $\chi(T)$ and
$\chi(H)$.  Closed diamonds correspond to the closing of the
spin-gap as determined from INS data in
figure~\ref{fig:phccprb2inelasfieldsweep}.  Closed squares represent
transitions to long range order as determined from the field and
temperature dependence of the $(\frac{1}{2} 0 \bar{\frac{1}{2}})$
antiferromagnetic Bragg peak shown in
figures~\ref{fig:phccprb2fieldsweeps} and
~\ref{fig:phccprb2tempsweeps}. (c) Log-log plot of the
low-temperature reentrant behavior of the LRO  phase.}
\end{figure}

A second view of the magnetic susceptibility data shown in figures
\ref{fig:phccprb2chiofhparb} and \ref{fig:phccprb2chioft} is
provided by the color contour map in figure
\ref{fig:phccprb2phasediag}(a).  Since  the sharp features in
$\chi(H)$ at the lowest-$T$ herald the  onset of finite
magnetization at $H_{c1}$ and the entry into the fully polarized
(FP) regime at $H_{c2}$, we identify the continuation of these
features for $T\leq 5$ K as experimental signatures of the closing
of the low-field
spin gap  and the opening of the high-field spin gap with increasing
field in the vicinity of the critical fields $H_{c1}$ and $H_{c2}$,
respectively. These features provide a starting point in
constructing an $H$-$T$ phase diagram for PHCC, and are shown as
open circles in figure~\ref{fig:phccprb2phasediag}(b).  By extending
these points through inclusion of the sharp features visible in
$\chi(T)$ for $T > 5$ K, they are seen to demarcate the entire
large-susceptibility region in figure
\ref{fig:phccprb2phasediag}(a).  At non-zero $T$, these points
delineate a crossover, rather than a phase transition.  Near the
lower QCP $H_{c1}$, for example, this boundary marks the change from
the non-magnetized quantum paramagnet (spin-gap) regime to a
quasi-2D renormalized classical (RC-2D) region with finite
magnetization.

We also identify two additional crossovers in the $H$-$T$ plane of
PHCC that are defined by rounded maxima in $\chi(T)$ for $H <
H_{c1}$ and $H > H_{c2}$.  These are demarcated by open diamonds in
figure~\ref{fig:phccprb2phasediag}(b).  At low field, the resulting
boundary indicates where antiferromagnetic correlations begin to
develop as the system is cooled toward the spin gap regime.  This
process has been well-studied in other spin gap systems, such as
Haldane spin-1 chains \cite{mahaldane}. The intersection of this
crossover with the zero-field axis in
figure~\ref{fig:phccprb2phasediag}(b) marks the onset of a
high-temperature semi-classical paramagnetic regime (P) where the
system has not yet attained significant antiferromagnetic
correlations and the magnetic susceptibility is that of a disordered
ensemble of exchange-coupled spins with one energy scale,
\textit{i.e.} Curie-Weiss behavior.

The high field crossover regime (open diamonds above $H = H_{c2}$ in
figure~\ref{fig:phccprb2phasediag}(b)) is not as well characterized
experimentally,
but should mark the onset of appreciable ferromagnetic correlations
as the system is cooled at high field toward the FP state. The
excitations in the FP state are expected to be ferromagnetic spin
waves.   As for a conventional ferromagnet in a field, these
excitations will show a field-dependent gap, although here the
energy scale for the gap should be set by $\Delta_{FP} = g\mu_B(H -
H_{c2})$. Fits of the form $\chi(T) \propto e^{-\Delta/k_B T}$, such
as those shown in figure~\ref{fig:phccprb2chioft}(c), yield
estimates for the gap in the FP regime.  Values of $\Delta_{FP}$ as
determined from a simultaneous fit of $\chi(T)$ in the temperature
range $0.75 \leq T \leq 6$ K for $40 \leq \mu_0 H \leq 47$ T are
shown in the inset of figure~\ref{fig:phccprb2chioft}(c) and
are consistent with a linear increase of $\Delta_{FP}$ with field
above $H_{c2}$.

The evolution of the magnetic susceptibility as a function of
magnetic field and temperature has been examined numerically for the
two-leg spin-ladder \cite{wangprl2000}.  Calculations depict a
low-temperature field dependent magnetization, which is symmetric
about $H_{m} = \frac{1}{2}(H_{c1} + H_{c2})$ with square root
singularities near the upper and lower critical fields, consistent
with  results for other 1D spin-gap systems \cite{chitraprb1997}.
Its phase diagram is thus similar to our findings for PHCC. The
calculations also report similar susceptibility anomalies:  a double
peaked function at low temperatures, a single rounded peak that
moves to lower fields as the temperature is increased, and a
monotonically decreasing field dependence as the temperature is
increased beyond a certain temperature, $T_{{\mathrm cl}}$.  A
crossover temperature marking the boundary between the classical and
quantum regimes, was identified as $T_{{\mathrm
cl}}=1.109\Delta/k_{B}$ for the two-leg spin ladder.  For PHCC, this
corresponds to $T_{{\mathrm cl}}= 1.109\Delta/k_{B} = 13.1$ K which
agrees well with the zero-field lower temperature bound of the
crossover regime defined in figure~\ref{fig:phccprb2phasediag} where
we find $T_{{\mathrm cl}} \approx 15.4$ K based upon susceptibility
measurements.

Close to the critical fields the low-temperature field dependent
magnetization should follow power laws of the form $M \propto
|H-H_{c}|^{1/\delta}$, where $H_{c}$ is a critical field
\cite{sakaiprb1998}.  In terms of magnetic susceptibility, this
power law becomes \numparts
\begin{eqnarray}
 \chi_{+} = dM/dH \propto (H-H_{c1})^{1/\delta_{1}-1}
 \label{eq:phccchieqns1} \\
 \chi_{-} = dM/dH \propto (H_{c2}-H)^{1/\delta_{2}-1}
 \label{eq:phccchieqns2}
\end{eqnarray}
\endnumparts
immediately above ($\chi_{+}$) and below ($\chi_{-}$) the respective
lower and upper critical fields.  For PHCC
Figures~\ref{fig:phccprb2chiofhparb} and
\ref{fig:phccprb2chithreeorient} indicate that $\delta_{2} >
\delta_{1} \approx 1$ to account for the observed curvature near the
critical fields. Because $\delta_{1} \approx 1$, we
fit $M(H)$ rather than $\chi(H)$ to determine $\delta$.  This is
done simultaneously for all $T=0.46$ K single crystal orientations
 with a common value of $\delta_{1}$ over
a range of one Tesla above $H_{c1}$. The single crystal
low-temperature, $T=0.46$ K data were also simultaneously fit over a
range of one Tesla below $H_{c2}$ using
Eqn.~(\ref{eq:phccchieqns2}). The simultaneous fits shared a common
$\delta$ parameter; whereas critical fields and overall amplitudes
were allowed to vary between the different orientations.  The
resulting values, $\delta_{1}=0.91(1)$ and $\delta_{2} = 1.8(2)$,
and their corresponding power-laws are shown in
figures~\ref{fig:phccprb2chiofhparb}
and~\ref{fig:phccprb2chithreeorient}. For comparison, 1D singlet
ground state systems are predicted to have symmetric susceptibility
and magnetization curves with  $\delta_{1}=\delta_{2}=2$
\cite{sakaiprb1998}, and the 2D square lattice Heisenberg model has
an analogous upper critical field with predictions ranging from
$\delta_{2}=1$ to $\delta_{2}=1$ with logarithmic corrections
\cite{zhitomirskyprb1998,yangmutterzphys}. An interpretation of
$H_{c1}$ as magnon Bose-Einstein condensation yields the prediction
of $\delta_1=1$ (see section \ref{becsec}).

\subsection{Specific heat}

We compare our specific heat measurements to the frustrated bilayer
model of Gu and Shen \cite{gueurphysb2000}.  Their model contains  a
strong interlayer bond $J_0$, a weaker intralayer bond $J_1$, and a
frustrating interlayer  bond $J_2$.  This exchange topology is
similar to that in PHCC.  We include a magnetic field dependent term
in their thermodynamic expansion via a linear Zeeman splitting of
the triplet excited state as a function of applied magnetic field.
We carried out  a global fit of all our specific heat data with
$\mu_0 H \leq 9$ T and $T \leq 30$ K using only the free energy
terms up to powers of $1/E^{-2}$ in Gu and Shen's expansion, one
independent prefactor for each magnetic field, and global terms
consisting of the three exchange constants, a $g$-factor and a
lattice contribution $\propto T^3$.  The results of this fit are
shown as solid lines in figure~\ref{fig:phccprb2cp1} and account
very well for the temperature and field dependence with $J_0 =
2.08(1)$ meV, $J_1 = 0.34(3)$ meV and $J_2 = -0.13(3)$ meV

\begin{figure}[t]
\centering\includegraphics[scale=0.80]{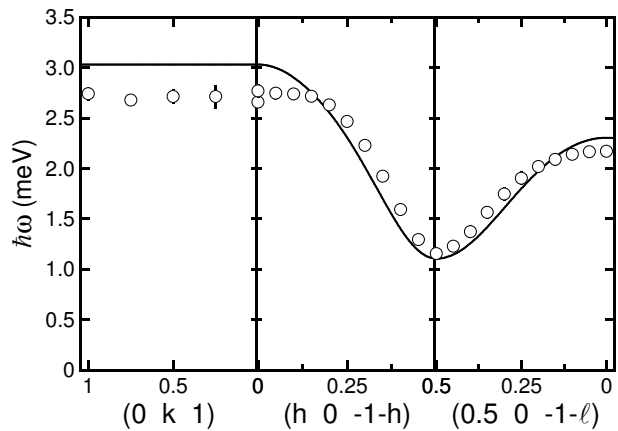}
\caption{\label{fig:phccprb2cp2} Zero-field magnon (or triplon)
dispersion in PHCC. Data taken from Refs~\cite{stonephccprb01} and
\cite{stonephccnature}. Solid lines are the calculated dispersion
based upon fitting magnetic heat capacity to the bilayer model as
described in the text.}
\end{figure}

The  exchange parameters for the frustrated bilayer model determined
from the specific heat data for PHCC can be used to evaluate the
corresponding dispersion relation, equation 4 of
Ref.~\cite{gueurphysb2000}.  Figure~\ref{fig:phccprb2cp2} depicts
the calculated dispersion for PHCC along several high-symmetry
directions in reciprocal space, compared to the  zero-field triplon
dispersion measured by inelastic neutron scattering
\cite{stonephccprb01,stonephccnature}.  The values derived from the
fits to $C_m$ for the spin gap $\Delta = 1.11(1)$ meV and bandwidth
of 1.92(2) meV  agree quite well with the measured spin gap, $\Delta
= 0.99$ meV and bandwidth, 1.73 meV.  Furthermore, the frustrated
bilayer model reproduces the shape of the dispersion remarkably well
given that specific heat data provided the only experimental input.

The lambda-type peaks in the magnetic heat capacity yield an
additional measure of the boundary on the magnetic field vs.
temperature phase diagram for PHCC,
figure~\ref{fig:phccprb2phasediag}(b), between the quasi-2D RC-2D
disordered phase and the LRO phase. However, $C_p$ was measured with
${\bf H}\parallel {\mathbf a}^{*}$, while all the other data
included in figure~\ref{fig:phccprb2phasediag}(b) were measured with
${\bf H}\parallel {\mathbf b}$.  As noted above, differences in the
critical fields measured for $\chi(H)$ for different field
orientations are accounted for by g-factor anisotropy.  After
rescaling by the ratio $H_{c1}({\bf H} \parallel {\mathbf b}) /
H_{c1}({\bf H} \parallel {\mathbf a}^{*}) = 0.917$ determined from
$\chi(H)$, the phase boundary derived from $C_p(H,T)$ is seen to
agree very well with that determined from elastic neutron scattering
as shown in figure~\ref{fig:phccprb2phasediag}(b) and (c) and
discussed below.

\begin{figure}[t]
\centering\includegraphics[scale=0.75]{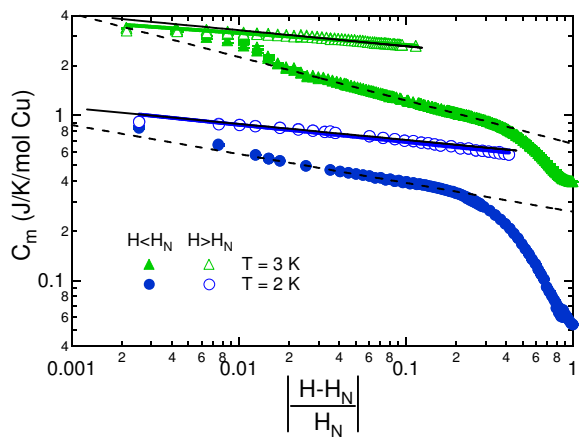}
\caption{\label{fig:phccprb2cp3} Magnetic heat capacity as a
function of reduced magnetic field, $h$, in the vicinity of the
phase transition to LRO. $H<H_{N}$ data (solid symbols) are plotted
against $h=\frac{H_N - H}{H_N}$.  $H>H_N$ data (open symbols) are
plotted against $h=\frac{H-H_N}{H_N}$.  Black solid lines correspond
to a global fit of $H>H_N$ data to a power law as discussed in text.
Coloured solid lines correspond to individual power law fits for
$H>H_N$. Individual fit results are very similar to the global fit
result. Dashed lines correspond to individual power law fits for
$H<H_N$ for $0.025 \leq h \leq 0.2$ in the RC-2D regime.}
\end{figure}

The magnetic field dependent specific heat shown in
figure~\ref{fig:phccprb2cp1}(c) also provides a sensitive
measure of critical scaling in the vicinity of the LRO transition.
$C_m(H)$ is shown in figure~\ref{fig:phccprb2cp3} as a function
of reduced magnetic field, $h = |\frac{H - H_N}{H_N}|$ near
the transition field $H_{N}$.  Power law behavior is clearly
seen for $H > H_N$, while below $H_N$, the magnetic specific
heat asymptotically approaches the same power law behavior as
the $H > H_N$ portion of the data.  In addition, there is a
region of reduced field for $H<H_N$, $0.02\leq h \leq 0.2$,
where we observe a different power-law behavior, \textit{i.e.}
a different slope on the log-log scale plot as indicated by the
dashed line in figure~\ref{fig:phccprb2cp3}.
The two different power-laws observed for $H<H_N$, as shown by
the fits for $0.02\leq h \leq 0.2$ and the asymptotic change in
power-law as one approaches $H_N$ is a potential
indication of an additional
disordered portion of the phase diagram for $H<H_N$.
This is in fact directly measured as a gapless disordered region
through comparisons of the magnetic susceptibility and neutron
scattering measurements as we discuss later in the text and as
shown in figures~\ref{fig:phccprb2phasediag}(b) and (c).

We fit the $H > H_N$ data in figure~\ref{fig:phccprb2cp3} to a
power law of the form
\begin{equation}
 C_m =  A(h)^{-\alpha},\label{eq:cppowerlaw}
\end{equation}
where $\alpha$ is the specific heat exponent. A global fit for $T=2$
K and $3$ K results in $\alpha =  0.097(2)$ while individual fits
yield $\alpha (T=2 \mathrm{K}) = 0.104(3)$ and $\alpha (T=3
\mathrm{K}) = 0.072(2)$.  Individual fits to Eq.~\ref{eq:cppowerlaw}
for $H<H_N$ over the range $0.02\leq h \leq 0.2$ yield $\alpha(T=2
\mathrm{K})=0.175(2)$ and $\alpha(T=3 \mathrm{K})=0.2626(7)$.  These
values are larger than those obtained for $H>H_N$, which indicates
that the fits include data beyond the critical regime.
\Tref{tab:exponenttable} summarizes theoretical and experimental
results for the 3D magnets.

\Table{\label{tab:exponenttable}Measured  exponents for PHCC
compared to calculated exponents listed in
\cite{collinsbook,campostriniexponents,xycalculatedref,campostriniexponent2}
for both 3D Heisenberg (HB), XY (3D), and Ising (2D and 3D)
transitions. Errors in referenced values are listed when available.
Error bars for $\alpha$ in PHCC correspond to one standard deviation
greater and less than the respective maximum and minimum fitted
values described in the text.} \br
& &  \centre{1}{Calc. HB} &\centre{1}{Calc. XY} & 2D Ising &\centre{1}{Calc. 3D Ising}\\
\ns
&  &\crule{1}  & \crule{1} & \crule{1} & \crule{1}\\
&PHCC&Ref~\cite{campostriniexponents}&Ref~\cite{xycalculatedref}&Ref~\cite{collinsbook}&Ref~\cite{campostriniexponent2}\\
\mr
$\alpha$ & $0.097^{+0.010}_{-0.027}$              & -0.1336(15)   & -0.0146(8)  & 0 & 0.1099(7)\\

$\beta$  & 0.34(2)            & 0.3689(3)$^*$        & 0.3485(2)$^*$ & 0.125 & 0.32648(18)$^*$\\

$\gamma$ & 1.22(4)$^*$         & 1.3960(9)         & 1.3177(5)   & 1.75 & 1.2372(4)  \\

$\delta$ & 4.6(2)$^*$      & 4.783(3)$^*$    & 4.780(2)$^*$ & 15 &4.7893(22)$^*$\\

$\nu$    & 0.634(1)$^*$        & 0.7112(5)      & 0.67155(27) & 1 &0.63002(23)  \\

$\eta$   & 0.07(6)$^*$     & 0.0375(5)       & 0.0380(4)  & 0.25 & 0.0364(4) \\

\br
\end{tabular}
\item[] $^{*}$Calculated from scaling relations and other listed critical exponents.
Error bars propagated from the measurements or calculations via the
scaling laws.
\end{indented}
\end{table}

\subsection{Elastic neutron scattering}
\subsubsection{Ordered magnetic structure}
To determine the field induced magnetic structure we measured
fourteen  magnetic Bragg peaks in the $(h0l)$
plane. The measurements were carried out at $T=1.65$ K and $\mu_0 H=13.7$
T which is within the ordered phase as established by heat capacity
measurements. \Tref{tab:peaktable} shows raw experimental results in the form of
wave vector integrated intensities, $I_m(\tau_m)$, for transverse
(rocking) scans. $I_m(\tau_m)$ is related to the magnetic structure
of the material as follows:
\begin{equation}
I_m(\tau_m)={\cal C}\int {\cal R}_{\tau}({\bf q}_{\perp})d{\bf
q}_{\perp} N_m\frac{(2\pi)^3}{v_m} {\cal I}_m(\tau_m).
\end{equation}
Here $\cal C$ is a normalization factor that depends on aspects of
the instrument such as the incident beam flux that are not
considered in the elastic resolution function ${\cal R}_{\tau}({\bf
q})$\cite{chesseraxe} which is sharply peaked around ${\bf q}=0$.
$N_m$ and $v_m$ are the number of magnetic unit cells and the
magnetic unit cell volume respectively. The magnetic Bragg
scattering intensity is given by
\begin{equation}
{\cal I}_m({\bf Q})=r_0^2(|{\bf \cal F}({\bf Q})|^2-|\hat{\bf
Q}\cdot{\cal F}({\bf Q})|^2),
\end{equation}
where $r_0=5.38$~fm,\cite{lovesy}
\begin{equation}
{\bf \cal F}({\bf Q})=\sum_{\bf d}\frac{g_{\bf d}}{2}F_{\bf d}({\bf
Q}))e^{-W_{\bf d}({\bf Q})}{\bf S}_{\bf d}
\end{equation}
is the magnetic structure factor, $g_{\bf d}$ is the Lande g-factor
for the magnetic ion on site $\bf d$ within the magnetic unit cell,
and $F_{\bf d}({\bf Q})$ is the magnetic form factor of the
Cu$^{2+}$ ion \cite{squires,lovesy}. The magnetic structure enters
through the expectation values for the spin vector, ${\bf S}_{\bf
d}$ in units of $\hbar$.

\Tref{tab:peaktable} compares the measured scattering intensity
$I_m(\tau_m)$ to the calculated intensity for the uniaxial magnetic
structure that best fits the data and is shown in
figure~\ref{fig:phccstructure}. The ordered moment lies within $\sim
7(4)^{\circ}$  of the {\bf a-c}-plane and the in-plane component is
within $\sim 6(4)^{\circ}$ of the {\bf a}-axis. For $T=1.65$ K and
$\mu_0 H=13.7$ T the sub-lattice magnetization is $g\mu_{B}\langle S
\rangle =0.33(3)g\mu_{B}$, determined by normalization to incoherent
scattering of the sample. Assuming antiferromagnetic interactions,
the ordered structure is consistent with the sign of the bond
energies determined on the basis of zero-field inelastic neutron
scattering.\cite{stonephccprb01}

\begin{table}[t]
\caption{Measured and calculated relative magnetic Bragg peak
scattering intensities at the indicated magnetic reciprocal lattice
points $\tau_m=h{\bf a}^*+l{\bf c}^*$. The parentheses indicate
statistical errors only.  The minimum in the crystallographic
reliability index $R=\sum |I_m^{obs}-I_m^{calc}| / \sum
I_m^{obs}=0.3$ is found for the magnetic structure depicted in
figure~\ref{fig:phccstructure} and described in the text.}
\label{tab:peaktable}
\begin{indented}
\item[]\begin{tabular}{cccc}
\br \multicolumn{1}{c}{$h$} & \multicolumn{1}{c}{$l$} &
\multicolumn{1}{c}{$I_m$ (obs)} & \multicolumn{1}{c}{$I_m$ (calc)}
\cr \mr 1   & -1/2  &  0.00(1)    &   0      \cr 1   & 1/2 & 0.00(1)
&   0 \cr -1   & -1/2  &  0.00(1) & 0 \cr 1/2 & -1 & 0.00(1)  &   0
\cr -1/2  & -1 & 0.00(1)    &   0 \cr 3/2 & 1/2 & 0.01(1) & 0.002
\cr 1/2 & -1/2  &  1.00(2)    & 1 \cr -1/2 & -1/2  & 0.21(1) & 0.234
\cr 3/2 & -1/2  & 0.16(1) & 0.333 \cr 1/2 & -3/2 & 0.11(1) & 0.250
\cr 5/2 & 1/2  & 0.03(1)  & 0.027 \cr 7/2 & 5/2 & 0.04(2) & 0.154
\cr \br
\end{tabular}
\end{indented}
\end{table}

\subsubsection{Phase diagram}
Figures~\ref{fig:phccprb2fieldsweeps}, \ref{fig:phccprb2tempsweeps}
and \ref{fig:reentrantplot}(c) compare the temperature and field
dependence of the magnetic Bragg intensity at
$Q=(\frac{1}{2}0\bar{\frac{1}{2}})$ to power laws
$I(H)\propto(H-H_N)^{2\beta}$ and $I(T)\propto(T_N-T)^{2\beta}$ in
order to extract the critical exponent, $\beta$. Rounding in the
vicinity of the phase transition was accounted for as a distribution
in the critical field. Possible origins of the 3.6\% distribution
width include inhomogeneities in the applied field and chemical
impurities. Least squares fitting included only data within 0.75 T
and 0.33 K of criticality. The corresponding fits appear as solid
lines in the figures.

Values of $(T_{N},H_N)$ extracted from the analysis appear as filled
squares on Figure~\ref{fig:phccprb2phasediag}. There is a clear
distinction between the onset of a state of high differential
susceptibility as determined from $\chi(H,T)$
 and the boundaries of the long range ordered phase
(LRO) extracted from neutron diffraction data. We denote the state
of the system between these boundaries as 2D renormalized classical.
Spin correlations in this phase were examined by quasi-elastic
neutron scattering measurements along the $(\frac{1}{2} 0 l)$ and
$(h 0 \bar{\frac{1}{2}})$ directions through the $(\frac{1}{2} 0
\bar{\frac{1}{2}})$ wave-vector at $T=3$ K and $\mu_0 H=9$ T.  No magnetic
Bragg peaks were encountered. Instead the broad peak centered at
${\bf Q}=(\frac{1}{2} 0 \bar{\frac{1}{2}})$ indicates short range
correlations with the same local structure as the long range order
that sets in at slightly higher fields.

\subsubsection{Critical exponents}
The order parameter critical exponent, $\beta$, determined for each
of the fits to the magnetic field and temperature dependent scans is
shown versus temperature in the inset to
figure~\ref{fig:phccprb2fieldsweeps}. For $T\geq 0.5$ K, the average
value is $\beta=0.34(2)$ which is indistinguishable from the
critical exponent for the 3D Heisenberg \cite{campostriniexponents},
Ising ($\beta = 0.326$), and XY ($\beta = 0.345$)
models\cite{collinsbook}. However, the positive specific heat
critical exponent, $\alpha$, uniquely identifies the phase
transition as being in the 3D Ising universality class. This is
consistent with expectations for a  phase transition to long range
magnetic order in space group $P\bar{1}$. We also calculate the
remaining critical exponents based upon the Rushbrooke , $\alpha +
2\beta + \gamma = 2$, Fisher, $\gamma = \nu (2 - \eta)$, Josephson,
$2-\alpha = \nu d$, and Widom, $\delta = 1 + \frac{\gamma}{\beta}$,
scaling laws, assuming $d=3$ dimensions
\cite{collinsbook,goldenfeld}. The values are again consistent only
with those of 3D Ising transitions as summarized in
~\tref{tab:exponenttable}. Recalling the exponents
$\delta_{1}=1.022(2)$ and $\delta_{2}=1.8(1)$ in
~\eref{eq:phccchieqns1} and ~\eref{eq:phccchieqns2} determined from
pulsed field magnetic susceptibility data, note that these values
are different from  $\delta$ derived from the scaling relations.
This is because $\delta_1$ and $\delta_2$ are associated with the
uniform susceptibility while $\delta$ in
table~\tref{tab:exponenttable} characterizes the response to a
staggered field conjugate to the N\'{e}el order parameter.

The apparent order parameter critical exponent increases as
temperature is reduced, such that $\beta = 0.40(1)$ for $T\leq 0.43$
K (see inset to figure~\ref{fig:phccprb2fieldsweeps}). A possible
interpretation of this behavior is that the
system enters a crossover regime associated with
the approach to quantum criticality.
Indeed, in the $T=0$ limit the system should be above its upper
critical dimension such that mean field behavior with a critical
exponent $\beta_{mf}=0.5$ should be expected.

The field sweeps of the ${\bf Q}=(\frac{1}{2} 0 \bar{\frac{1}{2}})$
magnetic Bragg peak intensity shown in
figure~\ref{fig:reentrantplot}(a)-(c), indicate a reentrant phase
transition.  The $H-T$ phase boundary obtained by fitting these data
(figure~\ref{fig:phccprb2phasediag}(b)) show a local minimum in the
phase boundary  between the LRO and QP phases. A possible origin of
this anomaly is coupling to non-magnetic degrees of freedom such as
nuclear spin or lattice distortions that are unimportant at higher
temperatures but become important sufficiently close to the quantum
critical point\cite{phccprl1}.

\subsection{Inelastic neutron scattering}
\subsubsection{Gapped Excitations}
Figure \ref{fig:phccprb2inelaslowt} shows the magnetic field and
temperature dependence of the excitation spectrum at the critical
wave vector ${\bf Q}=(\frac{1}{2} 0 \bar{\frac{1}{2}})$ for $0\leq H
< H_{c2}$. We fit each spectrum to a sum of Gaussian peaks to
account for the magnetic excitations with a constant background and
a field independent Gaussian+Lorentzian lineshape centered at zero
energy to account for incoherent elastic nuclear scattering.  The
field dependence of the peak locations are plotted for the $T=0.06$
K and $T=4.5$ K data in figure~\ref{fig:phccprb2inelaslowt}(b). The
data demonstrate the triplet nature of the excited state through its
Zeeman splitting. Above $H_{c1}$, the rate of change of the peak
position as a function of field for the higher energy peak is twice
that of the lower energy excitation, indicating a magnetized ground
state. We fit the low $T$ field dependent peak locations to the
spectrum of an antiferromagnetically coupled spin-1/2 pair with
singlet-triplet splitting $\Delta$ subject to a magnetic field
described by ${\cal H}_Z=-g\mu_B\mu_0H$. Results of this
two-parameter fit to the low-temperature peak locations are plotted
as solid lines in figure~\ref{fig:phccprb2inelaslowt}(b), and yield
values of $\Delta=1.021(2)$ meV, $g=2.33(1)$, and $\mu_0
H_{c1}=\Delta/(g\mu_B)=7.58(3)$ T, which are in agreement with
zero-field inelastic neutron scattering and bulk magnetization data.

\begin{figure}[t]
\centering\includegraphics[scale=0.6]{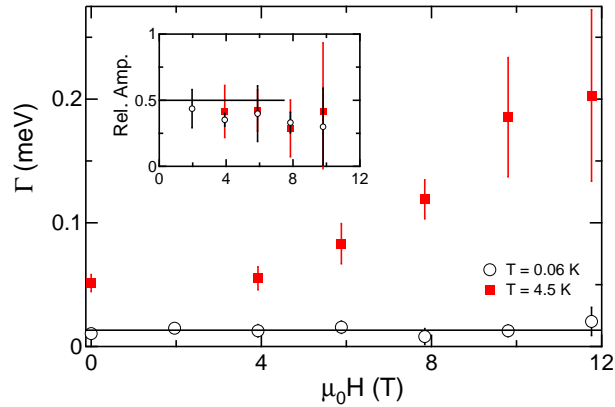}
\caption{\label{fig:phccprb2ivshwpeakchar}Characteristics of
constant energy scans as a function of applied magnetic field at
$T=0.06$ K (open symbols) and  $T=4.5$ K extracted through Gaussian
fits. Representative scans are shown in
figure~\protect\ref{fig:phccprb2inelaslowt}.  (a) Average resolution
corrected half width at half maximum for finite energy modes. Inset
shows the ratio of integrated intensities for the field
dependent modes to
the central field independent mode. The solid line indicates the
theoretical ratio of 0.5 expected for $H<H_{c1}$ at $T=0$.}
\end{figure}

The average resolution-corrected half width at half maximum,
$\Gamma$, of the three Gaussians that were fitted to each spectrum
for $T=0.065$ K and $T=4.5$ K are shown in
figure~\ref{fig:phccprb2ivshwpeakchar}. While $\Gamma$, considering
systematic uncertainties in the energy resolution, is
indistinguishable from zero at low $T$, it increases with applied
field at elevated temperature. Also increasing with field and
temperature is the magnon density. Hence the result is consistent
with an increased rate of magnon collisions as their density
increases. Self consistent mean field theories have been developed
to account for these effects.\cite{ruegg}

The inset to figure~\ref{fig:phccprb2ivshwpeakchar} shows
the field dependent ratio of the integrated intensities of the
central Gaussian peak  to that of the satellite peaks.  Based upon
the polarization factor and the Wigner-Eckart theorem as applied to
a Zeeman split triplet, this ratio should be 0.5 for $H<H_{c1}$
\cite{mullerprb1981}. Both the low- and high-temperature constant
wave-vector scans are consistent with this result. Above $H_{c1}$,
Bose condensation modifies the ground state and the experiment shows
a decrease of intensity for the highest energy mode.

\begin{figure}[t]
\centering\includegraphics[scale=0.6]{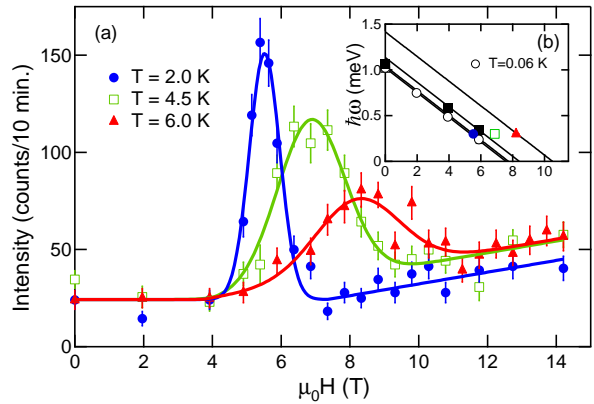}
\caption{\label{fig:phccprb2inelasfieldsweep} (a) Inelastic neutron
scattering intensity at $\hbar\omega = 0.3$ meV and $(\frac{1}{2} 0
\bar{\frac{1}{2}})$ as a function of applied magnetic field, ${\bf
H}\parallel {\mathbf b}$.  Solid lines are the result of a global
fit to a Gaussian peak with a sloping background above the magnetic
field $H_0$ and a constant background below the magnetic field
$H_0$.  Values of the constant background, $H_0$, and slope were
global parameters in the fit. (b) Energy transfer versus magnetic
field for peak locations as determined from (a) and the $T=4.5$ K
and $T=0.06$ K data in figure~\ref{fig:phccprb2inelaslowt}(a).
Symbols used in (b) correspond to those used in (a).  Solid lines
are fits to lines as described in the text.  The extrapolated
intercepts define an effective temperature dependent crossover
field.}
\end{figure}

We use inelastic neutron scattering to examine the crossover between
the low-field gapped (QP) and gapless (RC-2D) portions of the phase
diagram. The $T=0.06$~K lower critical field determined by comparing
the peak locations in low-temperature constant energy scans to
Eq.~\ref{eq:fandisp} is plotted in
figure~\ref{fig:phccprb2phasediag}(a) and (b) (solid diamond at $T =
0.06$ K).  With wave vector transfer at ${\mathbf Q} =(\frac{1}{2} 0
\bar{\frac{1}{2}})$ and energy transfer fixed at
$\hbar\omega=0.3$~meV the data in
figure~\ref{fig:phccprb2inelasfieldsweep} show the field dependent
intensity for various fixed temperatures.  At the lowest temperature
a sharp peak indicates the field at which the $S_z=1$ mode passes
through the 0.3 meV energy window of the detection system. With
increasing temperature and hence increasing magnon density not only
does the peak broaden as expected from an increased relaxation rate,
it also moves to higher fields indicating a finite slope for the
crossover line in the $H-T$ plane. Just as for the finite magnon
lifetime, the origin of this effect is magnon repulsion. The
increasing background is likely due to population of low-energy
gapless excitations which develop for magnetic fields above the
lower critical field. By extrapolation to the $\hbar\omega=0$ line
in Fig.~\ref{fig:phccprb2inelasfieldsweep}(b), these data were used
to determine the temperature dependence of the field required to
close the spin gap and the corresponding data points are included in
figure~\ref{fig:phccprb2phasediag}(a) and (b) (solid diamonds). The
coincidence between the crossover field determined in this manner
and the fields associated with differential susceptibility anomalies
establish a not unexpected association between effectively closing
the gap at the critical wave vector and maximizing the differential
uniform susceptibility.

\subsubsection{Magnetic Excitations in the high field paramagnetic state}
The previous results indicate a well defined crossover in PHCC from
a low field and low temperature state dominated by a gap in the
excitation spectrum to a high field state with an excitation
spectrum that has finite spectral weight at zero energy. The
crossover clearly occurs at lower fields than required to induce long
range magnetic order. There are possible analogies to certain
$S=1/2$ 1D Heisenberg antiferromagnetic systems including the linear
chain, the next nearest neighbor frustrated chain, and the
alternating chain where a low-energy continuum of excitations is
predicted to exist upon closure of the spin-gap
\cite{mullerprb1981,yuprb2000,hgaprb2001}. Calculations examining
both gapped and gapless $S=1/2$ spin-orbital chains describe an
incommensurate gapless continuum for large portions of exchange and
orbital interaction parameter space \cite{yu63prb2000}. A
characteristic of these continua are soft incommensurate modes
moving through the Brillouin zone from the antiferromagnetic zone
center to the ferromagnetic point as a function of applied magnetic
field \cite{chitraprb1997,tsvelik1995,batista2001,wang1992}.
Incommensurate magnetic field dependent soft modes are also found in
numerical calculations for the 2D antiferromagnetic square lattice
with next nearest neighbor interactions, a system with similar
antiferromagnetic interactions as PHCC \cite{yangmutterzphys}.

\begin{figure}[t]
\centering\includegraphics[width=8cm]{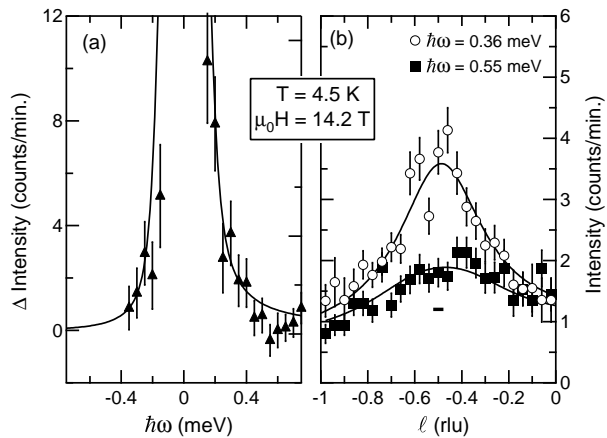}
\caption{\label{fig:phccprb2inelasqscans} (a)  Magnetic inelastic
scattering from PHCC at the critical wave vector $(\frac{1}{2} 0
\bar{\frac{1}{2}})$ for $\mu_0 H=14.2$~Tesla and $T=4.5$~K which is
immediately outside the long range ordered phase. A presumed
non-magnetic background measured at $(\frac{1}{2} 0 \bar{1})$ has
been subtracted. The solid line is a fit to a Lorentzian relaxation
function as described in the text. (b) Constant energy scans,
$\hbar\omega = 0.36$ meV, along the $(\frac{1}{2} 0 l)$ direction at
$T=4.5$ K and $\mu_0 H=14.2$ T for $\hbar\omega=0.36$ and $0.55$
meV. Solid lines through data are fits to a single peaked Lorentzian
functions with sloping backgrounds}
\end{figure}

An important distinction between PHCC and the systems mentioned
above lies in the fact that for PHCC the putative gapless state so
far was only observed at finite temperatures. Ascribing this to the
inevitable effects of inter-plane coupling, the analogue of which
occurs via interchain coupling in quasi-one-dimensional systems,
there exists a
potential for field induced incommensurate correlations in PHCC.
Here we present preliminary experiments at the highest fields
presently accessible to neutron scattering. The field was $\mu_0H=
14.2$~Tesla and the temperature chosen was $T=4.5$ K, which is 20\%
above the critical temperature at that field ($T_C(\mu_0H= 14.2$
T)=3.705(3)~K). Shown in Fig.~\ref{fig:phccprb2inelasqscans} (a) are
difference data indicating excess magnetic spectral weight at the
antiferromagnetic zone center $(\frac{1}{2} 0 \bar{\frac{1}{2}})$ as
compared to $(\frac{1}{2} 0 \bar{1})$. The solid line shows a fit to
the following resolution convolved Lorentzian relaxation spectrum,
\begin{equation}
S(\omega)\propto \frac
{1}{1-\exp(-\beta\hbar\omega)}\frac{\chi_0\Gamma\omega}{\Gamma^2+\omega^2},
\end{equation}
which obeys detailed balance and features a relaxation rate,
$\hbar\Gamma=0.025 \pm 0.2$~meV. The spectrum extends to
and is approximately
symmetric about zero energy, which is consistent with a 2D
renormalized classical phase close to the phase transition to
commensurate magnetic order.

Figure~\ref{fig:phccprb2inelasqscans} (b) shows two constant energy
scans at $\hbar\omega=0.36$~meV and $\hbar\omega=0.55$~meV that
indicate short range commensurate magnetic correlations. The fitted
Lorentzian peaks, which are much broader than the experimental
resolution (solid bar) correspond to a dynamic correlation length
scale of $3.2\pm2.5$~\AA~and $4.2\pm2.1$~\AA~for the
$\hbar\omega=0.55$~meV and $\hbar\omega=0.36$~meV measurement
respectively. The magnetization of PHCC is only 20\% of the
saturation magnetization in these measurements so any incommensurate
shift, which should be a similar fraction of the distance to the
Brillouin zone center is not clearly visible in the present data.

\subsection{Bose condensation of magnons}
\label{becsec}

In this section the phase transition to long range magnetic order in
PHCC is analyzed from the perspective of magnon condensation \cite{matsubara1956,affleck1990,Nikuni00,fisher1989}.
Following Nikuni et al. \cite{Nikuni00} and Misguich and Oshikawa
\cite{Misguich04} we treat magnons as bosons whose density $n$ is
proportional to the deviation of the longitudinal magnetization from
its plateau value $M_{z0}$: $g \mu_B n = |M_z-M_{z0}|$.  The
low-density magnon gas is described by the Hamiltonian
\begin{equation}
\mathcal{H} = \sum_\mathbf{k} (\epsilon_\mathbf{k}-\mu)
a^\dagger_\mathbf{k} a_\mathbf{k} + \frac{v_0}{V}
\sum_{\mathbf{q,k,k'}} a^\dagger_\mathbf{q+k'}
a^\dagger_\mathbf{q-k'} a_\mathbf{q-k} a_\mathbf{q+k},
\end{equation}
where $\mu = g \mu_B (H-H_{c1})$ is the boson chemical potential and
$v_0$ is a short-range repulsion.  The energy dispersion
$\epsilon_\mathbf{k}$ is assumed to have the global minimum at
$\mathbf{k}_0 = [\frac{1}{2}0\frac{1}{2}]$.  For convenience,
excitation energies are measured from the bottom of the band, so that
$\epsilon_{\mathbf{k}_0} = 0$.

At zero temperature the system acquires a condensate $\langle a_0
\rangle = \alpha \sqrt{V}$ when $\mu>0$.  The energy density of the
ground state in the presence of a condensate (neglecting quantum
corrections) is $v_0 |\alpha|^4 - \mu |\alpha|^2$, which is
minimized when $|\alpha|^2 = \mu/2v_0$.  In this approximation the
boson density is
\begin{equation}
n = |\alpha|^2 = g \mu_B (H-H_{c0})/2v_0. \label{eq-n}
\end{equation}
\noindent
$n$ is related to the longitudinal magnetization
\begin{equation}
M_z = g \mu_B n
\end{equation}

At a finite temperature we obtain an effective Hamiltonian
\begin{equation}
\mathcal{H} = \sum_\mathbf{k} (\epsilon_\mathbf{k}-\mu+v_0 n)
a^\dagger_\mathbf{k} a_\mathbf{k}, \label{eq-H}
\end{equation}
where $n$ is the density of thermally excited bosons.  The bosons
condense when the renormalized chemical potential $\tilde{\mu}
\equiv \mu - 2v_0 n$ vanishes.  This yields the critical density
\begin{equation}
n_c(T) = \frac{1}{V} \sum_\mathbf{k}
\frac{1}{\exp{({\epsilon_\mathbf{k}/T})}-1}. \label{eq-nc}
\end{equation}
The critical field is then
\begin{equation}
H_{c}(T) = H_{c}(0) + 2v_0 n_c(T)/g \mu_B. \label{eq-Hc}
\end{equation}
The boson scattering $v_0$ can be extracted from low-temperature
magnetization curves.  The magnon density per unit cell $n$ is equal
to the ratio $M/M_\mathrm{sat}$ (there is exactly one magnon per
unit cell when $M = M_\mathrm{sat}$).  Based on the magnetization
curve at the lowest available temperature $T = 0.46$~K we obtain
$v_0 = 1.91$ meV $\times$ the unit-cell volume (u.c.v.).  In other
words, if we model the magnons as bosons on a cubic lattice with
anisotropic hopping and on-site interactions $U$, the interaction
strength would be $U = 2$ meV. The linear behavior extends over a
range of some 10 T, or 13 K in temperature units.

\begin{figure}[t]
\centering\includegraphics[scale=1]{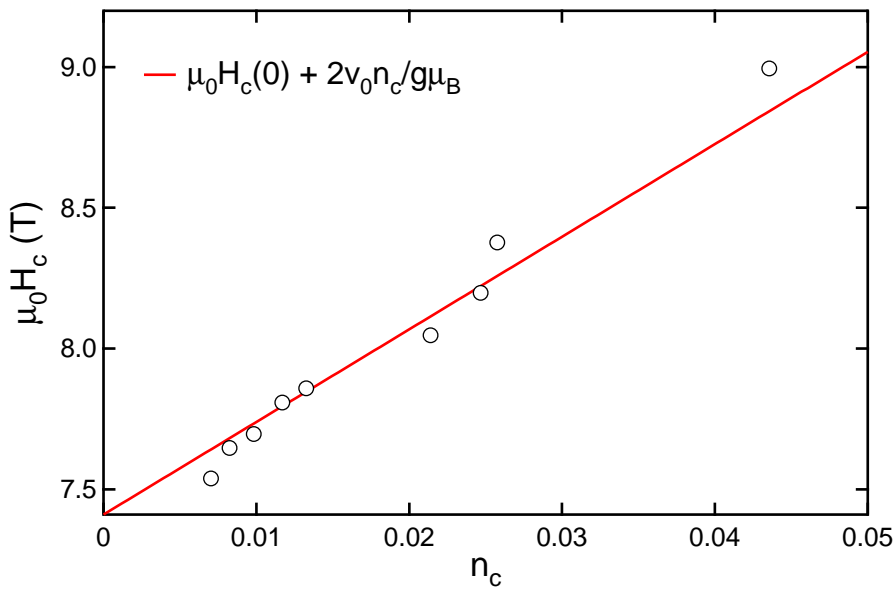}
\caption{Critical field $H_c$ versus critical magnon number per
unit cell $n_c$.  The straight line is Eq.~\ref{eq-Hc} with
$\mu_0 H_c(0) = 7.41$ T and $v_0 = 1.9$ meV u.c.v.}
\label{fig-Hc-vs-Mc}
\end{figure}

Using this value of the boson coupling we can test the predicted
relation between the critical field and critical magnon density
(\ref{eq-Hc}) at finite temperatures from the data for $n =
M/M_\mathrm{sat}$ in the temperature range 0.46 K $\leq T \leq 6$
K. From these we extracted pairs $(M,H)$ whose $H(T)$ values were
sufficiently close to the line of phase transitions $H_c(T)$.  The
resulting graph is shown in Fig.~\ref{fig-Hc-vs-Mc}.  The
zero-temperature critical field $H_{c1} = 7.41$ T was determined
from the onset of Neel order at the lowest temperatures ($T<0.3$ K).
The agreement with Eq.~(\ref{eq-Hc}) is fair.

The line $H_c(T)$ is obtained from Eqs.~(\ref{eq-nc}) and
(\ref{eq-Hc}).  The magnon dispersion is taken from Stone et al.
\cite{stonephccprb01} (the 6-term version).  That dispersion
relation $\epsilon_\mathbf{k}^\mathrm{2D}$ is strictly
two-dimensional, which precludes Bose condensation at a finite
temperature. We therefore added weak tunnelling between the planes:
\begin{equation}
\epsilon_\mathbf{k} = \epsilon_\mathbf{k}^\mathrm{2D} + 2\tau
[1-\cos{(k_y b)}].
\end{equation}
The experimental data \cite{Stone-thesis} give an upper bound for
the total bandwidth in the $[010]$ direction: $4\tau < 0.5$ meV.

When $T \ll \tau$, only the bottom of the magnon band has thermal
excitations and one may replace the exact band structure with a
parabolic dispersion to obtain
\begin{equation}
n_c(T) = \zeta(3/2) \, \left(\frac{m^\mathrm{3D} T}{2\pi
\hbar^2}\right)^{3/2},
\end{equation}
where $m^\mathrm{3D} = (m_a m_b m_c)^{1/3}$ is the 3D effective
mass. In the limit of a very weak interplane tunnelling we can also
access a regime where $\tau \ll T \ll W$, where $W \approx 2$ meV is
the magnon bandwidth.  In this regime the in-plane dispersion can
still be treated as parabolic.  This yields the critical density for
a quasi-2D Bose gas \cite{Micnas90}
\begin{equation}
n_c(T) = \frac{m^\mathrm{2D} T}{2\pi \hbar^2 b} \, \log{\frac{2
T}{\tau}},
\end{equation}
where $m^\mathrm{2D} = (m_a m_c)^{1/2}$ and $b$ is the interplane
distance.  In the intermediate regime $T = \mathcal{O}(\tau)$ the
critical density first rises {\em faster} than $T^{3/2}$
\cite{Misguich04} before crossing over to the $T \log{T}$ behavior.

\begin{figure}[t]
\centering\includegraphics[scale=1]{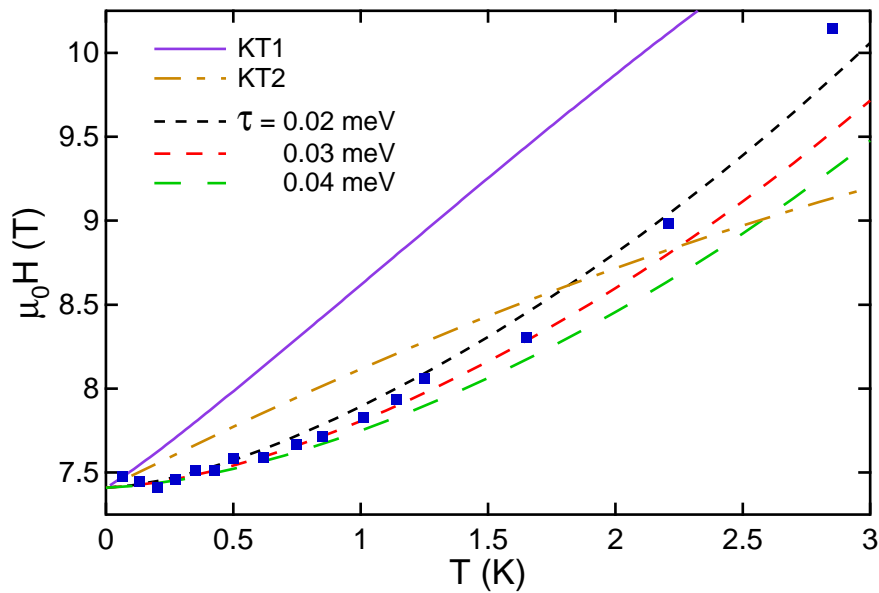}
\caption{Squares are data for $H_c(T)$ from elastic neutron
scattering.  $H_{c1}(T=0) = 7.41$ T, $v_0 = 1.91$ meV u.c.v.  Solid
lines are discussed in the text.}
\label{fig-Hc}
\end{figure}

Curves for $n_c(T)$ were obtained by integrating Eq.~(\ref{eq-nc})
numerically for several values of the interplane tunnelling $\tau$.
Since $v_0$ is already fixed, there are no further adjustable
parameters.  The data below 2.5 K ($H_c < 9$ T, $n_c < 0.05$ per
unit cell) are best described by the curves for $\tau = 0.03(1)$
meV. The results are shown in Fig.~\ref{fig-Hc}.

The theory based on the Hamiltonian (\ref{eq-H}) assumes that the
boson interaction $v_0$ is weak.  However, it is also applicable to
strongly interacting bosons, as long as the interaction is
short-range and the gas is sufficiently dilute.  In that case $v_0$
is not the bare interaction strength (which can be large) but rather
the scattering amplitude (the t-matrix) for a pair of low-energy
bosons.  For hard-core bosons on a cubic lattice with a
nearest-neighbor hopping amplitude $t$, one obtains $v_0 \approx 8t$
\cite{Friedberg94}.  In our case $t \approx W/8 = 0.25$ meV giving
$v_0 \approx 2$ meV. Another estimate along these lines, using the
``exact'' magnon dispersion yields, via
\begin{equation}
\frac{1}{v_0} = \frac{1}{V} \sum_\mathbf{k}
\frac{1}{2\epsilon_\mathbf{k}},
\end{equation}
yields a similar value of $v_0 = 2.5$ meV.  These numbers
demonstrate that the inferred interaction strength of 1.91 meV
u.c.v. is quite reasonable.

At higher temperatures the agreement worsens.  At $T=3.7$ K---the
highest temperature where the condensation is observed---the
mean-field theory predicts $H_c = 11$ T, well below the observed
value of 14.2 T.  The observed magnon density at the critical field
is 0.19 per unit cell (the theoretical estimate is 0.10).  It is
possible that the mean-field theory of magnon condensation,
appropriate for a dilute Bose gas of magnons, becomes unreliable due
to overcrowding.  (At a technical level, the assumption of pairwise
interactions may break down: two particles no longer collide in a
vacuum.)  Another possible reason for this failure is an increased
strength of (thermal) critical fluctuations.

\subsection{Kosterlitz-Thouless scenario}

If the planes in PHCC were fully decoupled from each other the Bose
condensation would change into the Kosterlitz-Thouless (KT)
vortex-unbinding transition.  It does not seem likely that the data
can be reconciled with this scenario: in $d=2$ spatial dimensions
the crossover exponent $\phi = 1$, so that the line of KT
transitions $H_c(T) = H_{c}(0) + C T^{\phi}$ is expected to be
linear.

The quantum criticality of the two-dimensional Bose gas was recently
examined by Sachdev and Dunkel \cite{Sachdev05} who used a
combination of scaling arguments and numerics to determine $H_c(T)$
for a weakly interacting Bose gas (\ref{eq-H}). (Analyzing the RG
flow alone does not solve the problem: the KT fixed point lies in a
strongly coupled limit.)

The RG analysis focuses on two variables: the (renormalized)
chemical potential $\tilde{\mu}$ and the two-particle scattering
amplitude $v$. Unlike in $d=3$ dimensions, the scattering amplitude
vanishes in the limit of zero energy.  Therefore the role of the
coupling constant is played by the t-matrix for particles with
energies $\mathcal{O}(T)$:
\begin{equation}
v = \frac{4\pi \hbar^2}{m \log{(\Lambda/T)}},
\end{equation}
where $\Lambda$ is a high-energy cutoff on the order of the magnon
bandwidth.

While both $\tilde{\mu}$ and $v$ flow under RG transformations, a
certain combination of these variables, namely $Tv/\tilde{\mu}$,
remains unchanged.  The value of this constant determines whether
the system flows to the disordered phase or to the phase with
long-range correlations.  The KT transition corresponds to
\begin{equation}
\frac{2mT v}{\hbar^2 \tilde{\mu}} = -K \approx -34.
\end{equation}
At the KT transition, the renormalized chemical potential is
\begin{equation}
\tilde{\mu} = -\frac{8\pi}{K \log{(\Lambda/T)}} T.
\end{equation}
We see that $|\tilde{\mu}|$ is a bit below $T$ (particularly if $T
\ll \Lambda$), which means that there are many thermally excited
particles.

The magnetic field is connected to the {\em bare} chemical
potential:
\begin{equation}
g\mu_B (H - H_{c0}) = \mu = \tilde{\mu} + 2 vn.
\end{equation}
The density is computed for an ideal gas:
\begin{equation}
n = \int \frac{d^2k}{(2\pi)^2} \frac{1}{e^{(\hbar^2
k^2/2m-\tilde{\mu})/T}-1} = \frac{m T}{2\pi \hbar^2}
\log{\frac{1}{1-e^{\tilde{\mu}/T}}} \approx \frac{m T}{2\pi \hbar^2}
\log{\frac{T}{-\tilde{\mu}}}.
\end{equation}
This yields the final result for $H_c(T)$:
\begin{equation}
g\mu_B (H - H_{c0}) = \mu = \frac{4 T}{\log{(\Lambda/T)}} \left[
\log{\left(\frac{K}{8\pi}\log{\frac{\Lambda}{T}}\right)} -
\frac{2\pi}{K} \right]. \label{eq-KT}
\end{equation}
It depends weakly on the ultraviolet cutoff $\Lambda$.  For $\Lambda
= 2.5$ meV, we have $\mu \approx 1.5 T$ in the temperature range of
interest.  The curve is labelled 'KT1' in Fig.~\ref{fig-Hc}.

The result (\ref{eq-KT}) can also be obtained directly from Eqs.~(3)
and (4) of Sachdev and Dunkel \cite{Sachdev05} if one takes the
classical limit $\mu \ll T$.  The value $\mu \approx 1.5 T$ suggests
that we are (just) outside the classical regime and one needs to
solve the Sachdev-Dunkel equations numerically.  The resulting curve
(for $\Lambda = 2.5$ meV) is shown in Fig.~\ref{fig-Hc} as 'KT2'.
It is also approximately linear, with a slope $d\mu/dT \approx 1$.

Even with the refined approach, the KT theory does not seem to agree
with the data.  As anticipated, it produces a linear dependence
$H_c(T)$ (up to logarithmic corrections) instead of a higher power
law.  The rise of the data above the KT line at high temperatures may
signal the inadequacy of the continuum approximation at high magnon
concentrations.

\section{Conclusions}
\label{conclusion}

Originally thought to be a weakly alternating spin chain, our
measurements establish PHCC as a quasi-two-dimensional spin-dimer
system. The gap is sufficiently small to allow access to the lower
critical field and the bandwidth sufficiently large to produce
several distinct phases.

The long range N\'{e}el ordered phase is a commensurate
antiferromagnet with moments perpendicular to the applied field
direction. The critical exponents for the phase transition are
similar to those of a 3D Ising model. The transition into this phase
can be described as magnon Bose-Einstein condensation. Mean-field
(Hartree-Fock) theory does an adequate job describing the
magnetization $M(H)$ at the lowest temperatures ($T = 0.46$ K) and
the shape of the critical line $H_c(T)$ for $T < 3$ K. The theory
predicts a small but nonzero magnon dispersion of $0.08 \geq 4\tau
\leq 0.16$ meV perpendicular to the 2D planes. The higher power of
the $H_c(T)$ dependence is caused by deviations of the magnon
density of states from the $E^{1/2}$ behavior of the continuum
theory.  The extra DOS (due to a van Hove singularity) is positive,
hence an increase over the $T^{3/2}$ law. The continuum KT theory
gives an approximately linear dependence $H_c(T)$ and thus does not
seem to do a good job.

Close to the lower quantum critical point there is an anomaly in the
phase boundary that is not accounted for by the BEC theory (see
figure~\ref{fig:phccprb2phasediag}(b) and
figure~\ref{fig:reentrantplot}(a)-(c).) Low T anomalies near the
zero temperature termination of phase boundaries are also seen
experimentally in the quantum liquids $^3$He\cite{pomeranchuk} and
$^4$He\cite{cordilloprb1998,stratyprl1966}, the frustrated spin
system GGG\cite{ramirezjapplphs1991,tsui1999}, and in numerical and
theoretical results for the BCC Ising antiferromagnet
\cite{landauprb1977}, the Ising antiferromagnet on a cubic lattice
\cite{huiprb1988}, and $S=1/2$ XXZ 1D- \cite{hieidaprb2001} and 2D-
\cite{schmidprl2002} antiferromagnets. In $^4$He the anomaly is
associated with there being more low energy degrees of freedom in
the solid than the liquid and a low$-T$ anomaly in the phase
boundary of LiHoF$_4$ has been associated with nuclear spin degrees
of freedom.\cite{bitko} For PHCC we speculate that features beyond a
spin model such as the relatively soft lattice or nuclear spins
become relevant sufficiently close to the quantum critical point.
Such a mechanism which appears to be rather general merits further
exploration through experiments that probe such auxiliary degrees of
freedom directly.

The high field paramagnetic phase is perhaps the most intriguing
aspect of PHCC. The crossover from the quantum paramagnet to this
phase remains remarkably sharp upon heating and the experiments show
that a finite temperature pseudo spin gap closes there. Further
analysis is required to understand why this crossover remains so
distinct. A proper examination of the gapless high field phase
requires inelastic neutron scattering magnetic fields of order 30
Tesla where the system is more heavily magnetized. This work must
await development of a higher field facility for inelastic neutron
scattering.

\ack We acknowledge discussions with A. Aharony, O. Entin-Wohlman,
A. B. Harris, S. Sachdev, and T. Yildirim. Work at JHU was supported by
the NSF through DMR-0074571, DMR-0306940, DMR-0348679 and
by the BSF through
grant No. 2000-073.  Work at PSU was supported by the NSF through
DMR-0401486.  ORNL is managed for the US DOE by UT-Battelle
Inc. under contract DE-AC05-00OR2272.  We thank P. Smeibidl and S.
Kausche for assistance with sample environment at HMI.  M. B. S.
thanks B. G. Ueland for assistance with heat capacity measurements
at PSU.

\end{document}